\pdfoutput=1

\documentclass[12pt,a4paper]{article}

\usepackage{multirow}
\usepackage{ifthen} 
\newboolean{pdflatex}
\setboolean{pdflatex}{true} 

\newboolean{articletitles}
\setboolean{articletitles}{true} 

\newboolean{uprightparticles}
\setboolean{uprightparticles}{false} 

\newboolean{inbibliography}
\setboolean{inbibliography}{false} 
\usepackage{longtable} 

\usepackage{dcolumn}   
\newcolumntype{.}{D{.}{.}{-1}}

\def\paperauthors{LHCb collaboration} 
\def\paperasciititle{Measurement of the B(s) --> D0bar K+ K-
      branching fractions} 
\def\papertitle{Observation of the decay $\Bs \to \Dzb \Kp\Km$ }
\def\paperkeywords{{High Energy Physics}, {LHCb}} 
\def\papercopyright{CERN on behalf of the LHCb collaboration}

\def\paperlicenceurl{https://creativecommons.org/licenses/by/4.0/}

\usepackage[top=1in, bottom=1.25in, left=1in, right=1in]{geometry}

\usepackage{microtype}
\usepackage{lineno}  
\usepackage{xspace} 
\usepackage{caption} 

\usepackage{graphicx}  
\usepackage{color}
\usepackage{colortbl}
\graphicspath{{./figs/}} 

\usepackage{amsmath} 
\usepackage{amssymb}
\usepackage{amsfonts}
\usepackage{upgreek} 

\newcommand*\patchAmsMathEnvironmentForLineno[1]{%
\expandafter\let\csname old#1\expandafter\endcsname\csname #1\endcsname
\expandafter\let\csname oldend#1\expandafter\endcsname\csname
end#1\endcsname
 \renewenvironment{#1}%
   {\linenomath\csname old#1\endcsname}%
   {\csname oldend#1\endcsname\endlinenomath}%
}
\newcommand*\patchBothAmsMathEnvironmentsForLineno[1]{%
  \patchAmsMathEnvironmentForLineno{#1}%
  \patchAmsMathEnvironmentForLineno{#1*}%
}
\AtBeginDocument{%
\patchBothAmsMathEnvironmentsForLineno{equation}%
\patchBothAmsMathEnvironmentsForLineno{align}%
\patchBothAmsMathEnvironmentsForLineno{flalign}%
\patchBothAmsMathEnvironmentsForLineno{alignat}%
\patchBothAmsMathEnvironmentsForLineno{gather}%
\patchBothAmsMathEnvironmentsForLineno{multline}%
\patchBothAmsMathEnvironmentsForLineno{eqnarray}%
}

\usepackage{hyperxmp}

\usepackage[pdftex,
            pdfauthor={\paperauthors},
            pdftitle={\paperasciititle},
            pdfkeywords={\paperkeywords},
            pdfcopyright={Copyright (C) \papercopyright},
            pdflicenseurl={\paperlicenceurl}]{hyperref}

\usepackage[all]{hypcap} 

\usepackage{xspace}
\usepackage{upgreek}

\def\lhcb {\mbox{LHCb}\xspace}

\def\babar  {\mbox{BaBar}\xspace}
\def\belle  {\mbox{Belle}\xspace}
\def\cleo   {\mbox{CLEO}\xspace}

\def\MagUp {\mbox{\em Mag\kern -0.05em Up}\xspace}

\ifthenelse{\boolean{uprightparticles}}%
{

 \def\Ppi         {\ensuremath{\uppi}\xspace}
 
 \def\Prho        {\ensuremath{\uprho}\xspace}

 \def\PDelta      {\ensuremath{\Delta}\xspace}
 \def\PXi      {\ensuremath{\Xi}\xspace}
 \def\PLambda      {\ensuremath{\Lambda}\xspace}
 \def\PSigma      {\ensuremath{\Sigma}\xspace}
 \def\POmega      {\ensuremath{\Omega}\xspace}
 \def\PUpsilon      {\ensuremath{\Upsilon}\xspace}

 \def\PB      {\ensuremath{\mathrm{B}}\xspace}
 
 \def\PD      {\ensuremath{\mathrm{D}}\xspace}

 \def\PK      {\ensuremath{\mathrm{K}}\xspace}

 \def\Pb      {\ensuremath{\mathrm{b}}\xspace}
 \def\Pc      {\ensuremath{\mathrm{c}}\xspace}

 \def\Pi      {\ensuremath{\mathrm{i}}\xspace}

 \def\Pp      {\ensuremath{\mathrm{p}}\xspace}

 \def\Ps      {\ensuremath{\mathrm{s}}\xspace}

}
{

 \def\Ppi         {\ensuremath{\pi}\xspace}
 
 \def\Prho        {\ensuremath{\rho}\xspace}

 \mathchardef\PDelta="7101
 \mathchardef\PXi="7104
 \mathchardef\PLambda="7103
 \mathchardef\PSigma="7106
 \mathchardef\POmega="710A
 \mathchardef\PUpsilon="7107
 
 \def\PB      {\ensuremath{B}\xspace}
 
 \def\PD      {\ensuremath{D}\xspace}

 \def\PK      {\ensuremath{K}\xspace}

 \def\Pb      {\ensuremath{b}\xspace}
 \def\Pc      {\ensuremath{c}\xspace}

 \def\Pi      {\ensuremath{i}\xspace}

 \def\Pp      {\ensuremath{p}\xspace}

 \def\Ps      {\ensuremath{s}\xspace}

}

\makeatletter
\ifcase \@ptsize \relax
  \newcommand{\miniscule}{\@setfontsize\miniscule{4}{5}}
\or
  \newcommand{\miniscule}{\@setfontsize\miniscule{5}{6}}
\or
  \newcommand{\miniscule}{\@setfontsize\miniscule{5}{6}}
\fi
\makeatother

\DeclareRobustCommand{\optbar}[1]{\shortstack{{\miniscule (\rule[.5ex]{1.25em}{.18mm})}
  \\ [-.7ex] $#1$}}

\def\squark    {{\ensuremath{\Ps}}\xspace}

\def\cquark    {{\ensuremath{\Pc}}\xspace}

\def\bquark    {{\ensuremath{\Pb}}\xspace}

\def\pion   {{\ensuremath{\Ppi}}\xspace}
\def\piz    {{\ensuremath{\pion^0}}\xspace}

\def\pip    {{\ensuremath{\pion^+}}\xspace}
\def\pim    {{\ensuremath{\pion^-}}\xspace}
\def\pipm   {{\ensuremath{\pion^\pm}}\xspace}

\def\rhomeson {{\ensuremath{\Prho}}\xspace}

\def\rhop     {{\ensuremath{\rhomeson^+}}\xspace}

\def\kaon    {{\ensuremath{\PK}}\xspace}
  \def\Kbar    {{\kern 0.2em\overline{\kern -0.2em \PK}{}}\xspace}

\def\KorKbar    {\kern 0.18em\optbar{\kern -0.18em K}{}\xspace}

\def\Kp      {{\ensuremath{\kaon^+}}\xspace}
\def\Km      {{\ensuremath{\kaon^-}}\xspace}
\def\Kpm     {{\ensuremath{\kaon^\pm}}\xspace}

\def\KS      {{\ensuremath{\kaon^0_{\mathrm{ \scriptscriptstyle S}}}}\xspace}

\def\Kstarz  {{\ensuremath{\kaon^{*0}}}\xspace}
\def\Kstarzb {{\ensuremath{\Kbar{}^{*0}}}\xspace}

\def\Kstarp  {{\ensuremath{\kaon^{*+}}}\xspace}

  \def\Dbar    {{\kern 0.2em\overline{\kern -0.2em \PD}{}}\xspace}
\def\D       {{\ensuremath{\PD}}\xspace}
\def\Db      {{\ensuremath{\Dbar}}\xspace}
\def\DorDbar    {\kern 0.18em\optbar{\kern -0.18em D}{}\xspace}
\def\Dz      {{\ensuremath{\D^0}}\xspace}
\def\Dzb     {{\ensuremath{\Dbar{}^0}}\xspace}

\def\Dstarz  {{\ensuremath{\D^{*0}}}\xspace}
\def\Dstarzb {{\ensuremath{\Dbar{}^{*0}}}\xspace}
\def\Dstarp  {{\ensuremath{\D^{*+}}}\xspace}
\def\Dstarm  {{\ensuremath{\D^{*-}}}\xspace}

\def\Dsp     {{\ensuremath{\D^+_\squark}}\xspace}
\def\Dsm     {{\ensuremath{\D^-_\squark}}\xspace}

\def\Dsmp    {{\ensuremath{\D^{\mp}_\squark}}\xspace}

\def\Dssmp   {{\ensuremath{\D^{*\mp}_\squark}}\xspace}

\def\B       {{\ensuremath{\PB}}\xspace}
\def\Bbar    {{\ensuremath{\kern 0.18em\overline{\kern -0.18em \PB}{}}}\xspace}

\def\BorBbar    {\kern 0.18em\optbar{\kern -0.18em B}{}\xspace}
\def\Bz      {{\ensuremath{\B^0}}\xspace}
\def\Bzb     {{\ensuremath{\Bbar{}^0}}\xspace}
\def\Bu      {{\ensuremath{\B^+}}\xspace}

\def\Bp      {{\ensuremath{\Bu}}\xspace}

\def\Bpm     {{\ensuremath{\B^\pm}}\xspace}

\def\Bd      {{\ensuremath{\B^0}}\xspace}
\def\Bs      {{\ensuremath{\B^0_\squark}}\xspace}
\def\Bsb     {{\ensuremath{\Bbar{}^0_\squark}}\xspace}

  \def\Y#1S{\ensuremath{\PUpsilon{(#1S)}}\xspace}

\def\proton      {{\ensuremath{\Pp}}\xspace}

\def\Xires       {{\ensuremath{\PXi}}\xspace}

\def\Lz          {{\ensuremath{\PLambda}}\xspace}
\def\Lbar        {{\ensuremath{\kern 0.1em\overline{\kern -0.1em\PLambda}}}\xspace}
\def\LorLbar    {\kern 0.18em\optbar{\kern -0.18em \PLambda}{}\xspace}

\def\Lb      {{\ensuremath{\Lz^0_\bquark}}\xspace}

\def\Lc      {{\ensuremath{\Lz^+_\cquark}}\xspace}

\def\Xibz    {{\ensuremath{\Xires^0_\bquark}}\xspace}

\def\BF         {{\ensuremath{\mathcal{B}}}\xspace}

\def\BR         {\BF}
\newcommand{\decay}[2]{\ensuremath{#1\!\to #2}\xspace}         

\def\to                 {\ensuremath{\rightarrow}\xspace}

\def\CP                {{\ensuremath{C\!P}}\xspace}

\def\AT#1     {\ensuremath{A_{\mathrm{T}}^{#1}}\xspace}           

\def\C#1      {\ensuremath{\mathcal{C}_{#1}}\xspace}                       
\def\Cp#1     {\ensuremath{\mathcal{C}_{#1}^{'}}\xspace}                    
\def\Ceff#1   {\ensuremath{\mathcal{C}_{#1}^{\mathrm{(eff)}}}\xspace}        
\def\Cpeff#1  {\ensuremath{\mathcal{C}_{#1}^{'\mathrm{(eff)}}}\xspace}       
\def\Ope#1    {\ensuremath{\mathcal{O}_{#1}}\xspace}                       
\def\Opep#1   {\ensuremath{\mathcal{O}_{#1}^{'}}\xspace}                    

\newcommand{\tev}{\ifthenelse{\boolean{inbibliography}}{\ensuremath{~T\kern -0.05em eV}}{\ensuremath{\mathrm{\,Te\kern -0.1em V}}}\xspace}
\newcommand{\gev}{\ensuremath{\mathrm{\,Ge\kern -0.1em V}}\xspace}
\newcommand{\mev}{\ensuremath{\mathrm{\,Me\kern -0.1em V}}\xspace}
\newcommand{\kev}{\ensuremath{\mathrm{\,ke\kern -0.1em V}}\xspace}
\newcommand{\ev}{\ensuremath{\mathrm{\,e\kern -0.1em V}}\xspace}
\newcommand{\gevc}{\ensuremath{{\mathrm{\,Ge\kern -0.1em V\!/}c}}\xspace}
\newcommand{\mevc}{\ensuremath{{\mathrm{\,Me\kern -0.1em V\!/}c}}\xspace}
\newcommand{\gevcc}{\ensuremath{{\mathrm{\,Ge\kern -0.1em V\!/}c^2}}\xspace}
\newcommand{\gevgevcccc}{\ensuremath{{\mathrm{\,Ge\kern -0.1em V^2\!/}c^4}}\xspace}
\newcommand{\mevcc}{\ensuremath{{\mathrm{\,Me\kern -0.1em V\!/}c^2}}\xspace}

\def\mum  {\ensuremath{{\,\upmu\mathrm{m}}}\xspace}

\def\gsim{{~\raise.15em\hbox{$>$}\kern-.85em
          \lower.35em\hbox{$\sim$}~}\xspace}
\def\lsim{{~\raise.15em\hbox{$<$}\kern-.85em
          \lower.35em\hbox{$\sim$}~}\xspace}

\def\ptot       {\mbox{$p$}\xspace}
\def\pt         {\mbox{$p_{\mathrm{ T}}$}\xspace}

\def\evtgen     {\mbox{\textsc{EvtGen}}\xspace}

\def\geant      {\mbox{\textsc{Geant4}}\xspace}

\def\photos     {\mbox{\textsc{Photos}}\xspace}

\def\pythia     {\mbox{\textsc{Pythia}}\xspace}

\def\tell1  {TELL1\xspace}
\def\ukl1   {UKL1\xspace}

\newcommand{\eg}{\mbox{\itshape e.g.}\xspace}
\newcommand{\ie}{\mbox{\itshape i.e.}\xspace}

\usepackage{cite} 
\usepackage{mciteplus}

\usepackage{longtable} 

\begin{document}

\renewcommand{\thefootnote}{\fnsymbol{footnote}}
\setcounter{footnote}{1}

\begin{titlepage}
\pagenumbering{roman}

\vspace*{-1.5cm}
\centerline{\large EUROPEAN ORGANIZATION FOR NUCLEAR RESEARCH (CERN)}
\vspace*{1.5cm}
\noindent
\begin{tabular*}{\linewidth}{lc@{\extracolsep{\fill}}r@{\extracolsep{0pt}}}
\ifthenelse{\boolean{pdflatex}}
 & & CERN-EP-2018-157 \\  
 & & LHCb-PAPER-2018-014 \\  
 & & July 5, 2018 \\ 
 & & \\
\end{tabular*}

\vspace*{3.0cm}

{\normalfont\bfseries\boldmath\huge
\begin{center}
  \papertitle
\end{center}
}

\vspace*{0.4cm}

\begin{center}
\paperauthors\footnote{Authors are listed at the end of this paper.}
\end{center}

\vspace{\fill}

\begin{abstract}
\vspace*{0.2cm}
  \noindent
The first observation of the $\Bs \to \Dzb \Kp\Km$ decay is reported, together with the most precise branching fraction measurement of the mode ${\Bz\to\Dzb\Kp\Km}$.  The results are obtained from an analysis of $pp$  collision data corresponding to an integrated luminosity of $3.0~\textrm{fb}^{-1}$. The data were collected with the \lhcb detector at centre-of-mass energies of $7$~and $8$~\tev.  The  branching fraction of the  ${\Bz\to\Dzb\Kp\Km}$ decay is measured relative to that of the decay $\Bz\to\Dzb \pip\pim$ to be
$$\frac{\BR(\Bz\to\Dzb\Kp\Km)}{\BR(\Bz\to\Dzb \pip\pim)} = (6.9 \pm 0.4 \pm 0.3)\%,$$
where the first uncertainty is statistical and the second is systematic. The measured branching fraction of the $\Bs \to \Dzb \Kp\Km$ decay mode relative to that of the corresponding
$\Bz$ decay is
$$\frac{\BR(\Bs\to\Dzb\Kp\Km)}{\BR(\Bz\to\Dzb\Kp\Km)} = (93.0 \pm 8.9 \pm 6.9)\%.$$
Using the known branching fraction of  ${\Bz\to\Dzb \pip\pim}$, the values
of ${{\cal B}\left(\Bd \to \Dzb \Kp\Km\right)=(6.1 \pm 0.4 \pm 0.3 \pm 0.3) \times 10^{-5}}$, and
${{\cal B}\left(\Bs \to \Dzb \Kp\Km\right)=}$ $(5.7 \pm 0.5 \pm 0.4 \pm 0.5) \times 10^{-5}$
are obtained, where the third uncertainties arise from the branching fraction of the decay modes ${\Bz\to\Dzb \pip\pim}$ and $\Bz\to\Dzb\Kp\Km$, respectively.

\end{abstract}

\vspace*{1.0cm}

\begin{center}
  Published in Phys.~Rev.~D98 (2018) 072006
\end{center}

\vspace{\fill}

{\footnotesize
\centerline{\copyright~2018 CERN for the benefit of the \lhcb collaboration. \href{http://creativecommons.org/licenses/by/4.0/}{CC-BY-4.0} licence.}}
\vspace*{2mm}

\end{titlepage}

\newpage
\setcounter{page}{2}
\mbox{~}

\cleardoublepage


\newcommand{\al}{\ensuremath{\kern 0.5em }}
\newcommand{\alll}{\ensuremath{\kern 0.35em }}
\newcommand{\all}{\ensuremath{\kern 0.25em }}

\newcommand{\alm}{\ensuremath{\kern -0.5em }}
\newcommand{\allm}{\ensuremath{\kern -0.25em }}
\renewcommand{\thefootnote}{\arabic{footnote}}
\setcounter{footnote}{0}

\newcommand{\extrn}{\ensuremath{\mathrm{\,(ext)}}\xspace}



\pagestyle{plain} 
\setcounter{page}{1}
\pagenumbering{arabic}


\section{Introduction}
\label{sec:Introduction}

The precise measurement of the angle $\gamma$ of the Cabibbo-Kobayashi-Maskawa (CKM) Unitarity Triangle~\cite{PhysRevLett.10.531,PTP.49.652}  is a central topic in flavour physics experiments. Its determination at the subdegree level in tree-level open-charm $b$-hadron decays is theoretically clean~\cite{Brod:2013sga,Brod:2014bfa} and provides a standard candle for measurements sensitive to new physics effects~\cite{CKMfitter2013}. In addition to the results from the  $B$ factories~\cite{Bevan:2014iga}, various measurements from \lhcb~\cite{HFLAV16,LHCb-PAPER-2016-032,LHCb-CONF-2018-002} allow  the angle $\gamma$ to be determined with an uncertainty of around $5^\circ$. However, no single measurement dominates the world average, as the most accurate measurements have an accuracy of about $10^\circ$ to $20^\circ$~\cite{LHCb-PAPER-2017-047,LHCb-PAPER-2018-017}. Alternative methods are therefore important  to improve the precision.  Among them, an analysis of the decay $\Bs \to \Dzb\phi$ has the  potential to make a significant impact~\cite{Gronau:1990ra,Gronau:2004gt,Gronau:2007bh,Ricciardi:1243068}. Moreover, a Dalitz plot analysis of $\Bs\to\Dzb\Kp\Km$ decays can further improve the determination of $\gamma$ due to the increased sensitivity to interference effects, as well as allowing the \CP-violating phase $\phi_s$ to be determined in $\Bs-\Bsb$ mixing  with minimal theoretical uncertainties~\cite{Nandi:2011uw}.

\begin{figure}[b]
\centering
\includegraphics[scale=0.60 ]{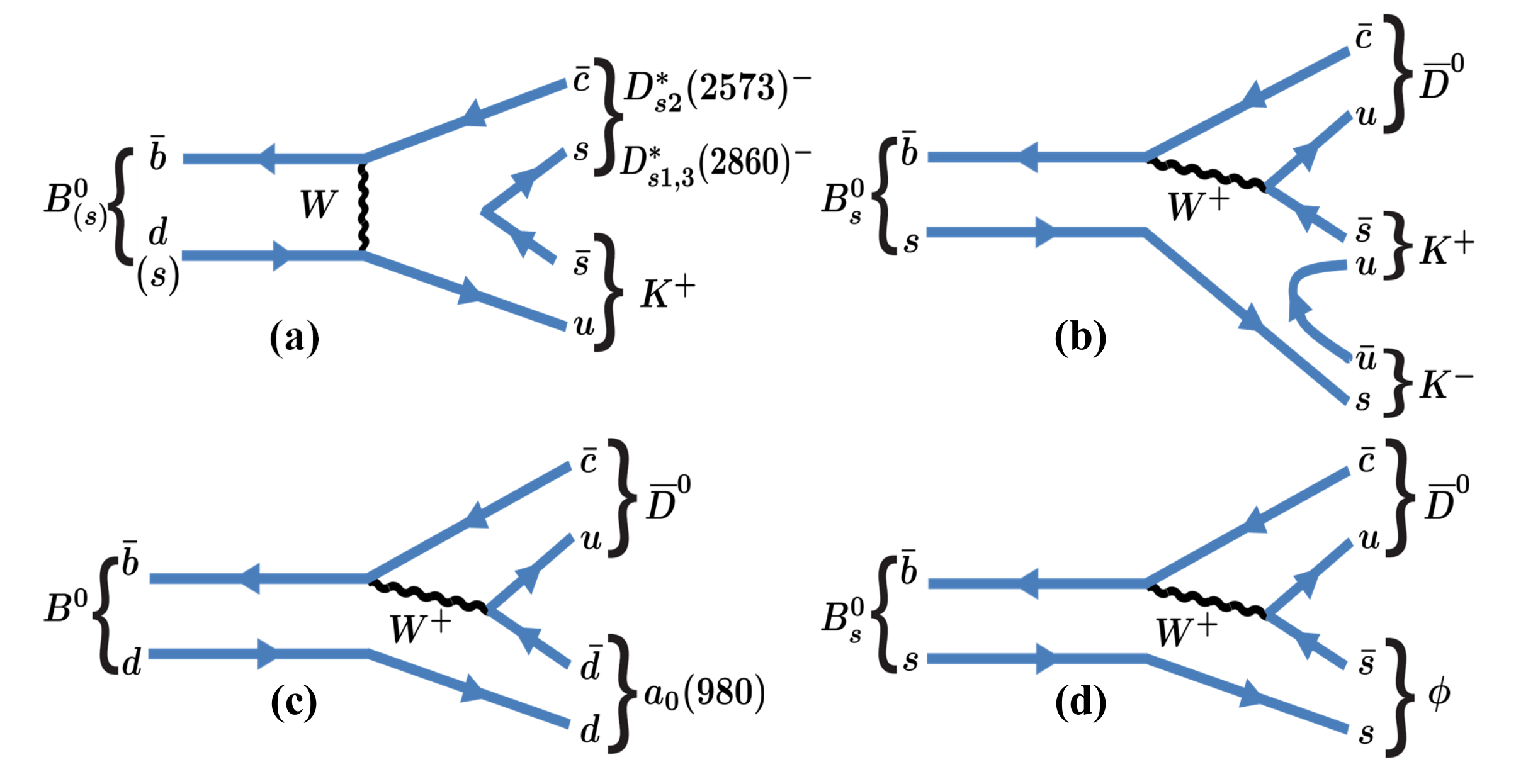}
\caption{Example Feynman diagrams that contribute to the ${B^0_{(s)} \to \Dzb\Kp\Km}$ decays via (a) $W$-exchange, (b) non-resonant three body mode,  (c) and (d) rescattering from a colour-suppressed decay.}
\label{fig:Feynman}
\end{figure}

The mode $\Bs \to \Dzb \phi$ has been previously observed by the \lhcb collaboration with a data sample corresponding to an integrated luminosity of $1.0~\mathrm{fb}^{-1}$~\cite{LHCb-PAPER-2013-035}. The observation of ${\Bd \to \Dzb \Kp\Km}$ and evidence for $\Bs \to \Dzb K^+K^-$ have also been reported by the \lhcb collaboration  using a data sample corresponding to $0.62~\mathrm{fb}^{-1}$~\cite{LHCb-PAPER-2012-018}. These decays are mediated by decay processes such as those shown in Fig.~\ref{fig:Feynman}.

In this paper an improved measurement of the branching fraction of the decay ${\Bd \to\Dzb\Kp\Km}$ and the first observation of the decay $\Bs\to\Dzb\Kp\Km$ are presented.\footnote{The inclusion of charge conjugate modes is implied throughout this paper.} The branching fractions are measured relative to that of the topologically similar and abundant decay $\Bz\to\Dzb \pip\pim$. The analysis is based on a data sample corresponding to an integrated luminosity of $3.0  \,{\rm fb}^{-1}$ of $pp$ collisions collected with the \lhcb~\mbox{detector}.
Approximately one third of the data was obtained during 2011, when the collision centre-of-mass energy was $\sqrt{s} = 7 \tev$, and the rest during 2012 with $\sqrt{s} = 8 \tev$.  Compared to the previous analysis~\cite{LHCb-PAPER-2012-018}, a revisited selection and a more sophisticated treatment of the various background sources are employed, as well as improvements in the handling of reconstruction and trigger efficiencies, leading to an overall reduction of  systematic uncertainties. The present analysis benefits from the improved knowledge of the decays $B^0_{(s)} \to \Dzb \Km\pip$~\cite{LHCb-PAPER-2013-022}, $\Lb \to \Dz p h^-$, where $h^-$  stands for a $\pim$ or a $\Km$ meson~\cite{LHCb-PAPER-2013-056}, which contribute to the background, and of the normalisation decay mode ${\Bd \to\Dzb\pip\pim}$~\cite{LHCb-PAPER-2014-070}.

This analysis sets the foundation for the study of the $B^0_{(s)} \to  \Db^{(*)0}\phi$ decays, which are presented in a separate publication~\cite{LHCb-PAPER-2018-015}.  The current data set  does not yet allow a Dalitz plot analysis of the $B^0_{(s)} \to \Dzb \Kp\Km$ decays to be performed, but these modes could provide interesting input to excited $\Dsp$ meson spectroscopy, in particular because the decay diagrams are different from those of  the $\Bs\to \Dzb \Km\pip$ decay~\cite{LHCb-PAPER-2014-036} (\ie\ different resonances can be favoured in each decay mode).

This paper is structured as follows.   A brief description of the LHCb detector, as well as the reconstruction and simulation software, is given in Sect.~\ref{sec:Detector}. Signal selection and background suppression strategies are summarised in Sect.~\ref{sec:Selection}. The characterisation of the various remaining backgrounds and their modelling is described in Sect.~\ref{sec:BkgModel} and the fit to the ${\Bz \to \Dzb\pip\pim}$  and ${B^0_{(s)} \to \Dzb\Kp\Km}$ invariant-mass distributions to determine the signal yields is presented in Sect.~\ref{sec:Fit}.  The computation of the efficiencies needed to derive the branching fractions is explained in Sect.~\ref{sec:Efficiency} and the evaluation of systematic uncertainties is described in Sect.~\ref{sec:Systematic}. The results on the branching fractions and a discussion of the Dalitz plot distributions are reported  in Sect.~\ref{sec:Results}.

\section{Detector and simulation}
\label{sec:Detector}

The \lhcb detector~\cite{Alves:2008zz,LHCb-DP-2014-002} is a single-arm forward spectrometer covering the \mbox{pseudorapidity} range $2<\eta <5$, designed for the study of particles containing \bquark or \cquark quarks. The detector includes a high-precision tracking system consisting of a silicon-strip vertex detector surrounding the $pp$ interaction region~\cite{LHCb-DP-2014-001}, a large-area silicon-strip detector located upstream of a dipole magnet with a bending power of about $4{\mathrm{\,Tm}}$, and three stations of silicon-strip detectors and straw drift tubes~\cite{LHCb-DP-2013-003} placed downstream of the magnet. The tracking system provides a measurement of momentum, \ptot, of charged particles with a relative uncertainty that varies from 0.5\% at low momentum to 1.0\% at 200\gevc. The minimum distance of a track to a primary vertex (PV), the impact parameter (IP), is measured with a resolution of $(15+29/\pt)\mum$, where \pt is the component of the momentum transverse to the beam, in\,\gevc. Different types of charged hadrons are distinguished using information from two ring-imaging Cherenkov (RICH) detectors~\cite{LHCb-DP-2012-003}. Photons, electrons and hadrons are identified by a calorimeter system consisting of scintillating-pad and preshower detectors, an electromagnetic calorimeter and a hadronic calorimeter. Muons are identified by a system composed of alternating layers of iron and multiwire proportional chambers~\cite{LHCb-DP-2012-002}.

The online event selection is performed by a trigger, which consists of a hardware stage, based on information from the calorimeter and muon systems, followed by a software stage, which applies a full event reconstruction. At the hardware trigger stage, events are required to have a muon with high \pt or a hadron, photon or electron with high transverse energy in the calorimeters. For hadrons, the transverse energy threshold is $3.5$~\gev. A global hardware trigger decision is ascribed to the reconstructed candidate, the rest of the event or a combination of both; events triggered as such are defined respectively as triggered on signal (TOS), triggered independently of signal (TIS), and triggered on both. The software trigger requires a two-, three- or four-track secondary vertex with a significant displacement from the primary $pp$ interaction vertices. At least one charged particle must have a transverse momentum ${\pt > 1.7 \gevc}$ and be inconsistent with originating from a PV. A multivariate algorithm~\cite{BBDT} is used for the identification of secondary vertices consistent with the decay of a \bquark hadron.

Candidates that are consistent with the decay chain $B^0_{(s)} \to \Dzb \Kp\Km$, $\Dzb \to \Kp\pim$ are selected. In order to reduce systematic uncertainties in the measurement, the topologically similar decay $\Bd\to\Dzb \pip\pim$, which has previously been studied precisely~\cite{Kuzmin:2006mw,LHCb-PAPER-2014-070}, is used as a normalisation channel. Tracks are required to be consistent with either the kaon or pion hypothesis, as appropriate, based on particle identification (PID) information from the RICH detectors. All other selection criteria are tuned on the $\Bz \to \Dzb \pip\pim$ channel. The large yields available in the normalisation sample allow the selection to be based on data. Simulated samples, generated uniformly over the Dalitz plot, are used to evaluate efficiencies and characterise the detector response for signal and background decays. In the simulation, $pp$ collisions are generated using \pythia~\cite{Sjostrand:2007gs} with a specific \lhcb configuration~\cite{LHCb-PROC-2010-056}.  Decays of hadronic particles are described by \evtgen~\cite{Lange:2001uf}, in which final-state radiation is generated using \photos~\cite{Golonka:2005pn}. The interaction of the generated particles with the detector, and its response, are implemented using the \geant toolkit~\cite{Allison:2006ve, *Agostinelli:2002hh} as described in Ref.~\cite{LHCb-PROC-2011-006}.

\section{Selection criteria and rejection of backgrounds}
\label{sec:Selection}

\subsection{Initial selection}
\label{sec:Prefilter}
Signal $B^0_{(s)}$ candidates are formed by combining \Dzb candidates, reconstructed in the decay channel $\Kp\pim$, with two additional tracks of opposite charge. After the trigger, an initial selection, based on kinematic and topological variables, is applied to reduce the combinatorial background by more than two orders of magnitude. This selection is designed using simulated ${\Bz \to \Dzb \pip \pim}$ decays as a proxy for signal and data ${\Bz \to \Dzb \pip \pim}$ candidates lying in the upper-mass sideband $[5400, 5600]$~\mevcc as a background sample. The combinatorial background arises from random combinations of tracks that do not come from a single decay. For the ${\Bz \to \Dzb \pip\pim}$ mode, no $b$-hadron decay contribution is expected in the upper sideband $[5320,6000]$~\mevcc, \ie\ no $\Bs$  contribution is expected~\cite{LHCb-PAPER-2012-056}.

The reconstructed tracks are required to be inconsistent with originating from any PV. The \Dzb decay products are required to originate from a  common vertex with an invariant mass within $\pm 25$~\mevcc of the known \Dzb mass~\cite{PDG2018}. The invariant-mass resolution of the reconstructed $\Dzb$ mesons is about $8$~\mevcc and the chosen invariant-mass range allows most of the background from the ${\Dzb \to \Kp\Km}$ and ${\Dzb \to \pip\pim}$ decays to be rejected.   The \Dzb candidates and the two additional tracks are required to form a vertex. The reconstructed \Dzb and \Bz vertices must  be significantly displaced from the associated PV, defined, in case of more than one PV in the event, as that which has the smallest $\chi^2_{\rm IP}$ with respect to the $B$ candidate. The $\chi^2_{\rm IP}$ is defined as the difference in the vertex-fit quality $\chi^2$ of a given PV reconstructed with and without the particle under consideration. The reconstructed \Dzb vertex is required to be displaced downstream from the reconstructed  $B^0_{(s)}$  vertex, along the beam axis direction. This requirement reduces the background from charmless $B$ decays, corresponding to genuine ${\Bz\to\Kp\pim h^+h^-}$ decays, for instance from ${\Bz\to \Kp\pim\rho^0}$ or ${\Bz\to \Kstarz \phi}$ decays, to a negligible level. This requirement also suppresses background from prompt charm production, as well as fake reconstructed $\Dzb$ coming from the PV. The $B^0_{(s)}$ momentum vector and the vector connecting the PV to the $B^0_{(s)}$ vertex  are requested to be aligned.

Unless stated otherwise, a kinematic fit~\cite{Hulsbergen:2005pu} is used to improve the invariant-mass resolution of the $B^0_{(s)}$ candidate.  In this fit, the $B^0_{(s)}$ momentum is constrained to point back to the PV and the \Dzb-candidate invariant mass to be equal to its known value~\cite{PDG2018}, and the charged tracks are assigned the $K$ or $\pi$ mass hypothesis as appropriate. Only ${B^0_{(s)} \to \Dzb h^+h^-}$ candidates with an invariant mass ($m_{\Dzb h^+h^-}$) within the range $[5115,6000]$~\mevcc are then considered. This range allows the $B^0_{(s)}$ signal regions to be studied, while retaining a sufficiently large upper sideband to accurately determine the invariant-mass shape of the surviving combinatorial background. The lower-mass limit removes a large part of the complicated partially reconstructed backgrounds and has a negligible impact on the determination of the signal yields.

The world-average value of the branching fraction ${\BR(\Bz \to \Dzb \pip\pim)}$ is equal to  $(8.8 \pm 0.5) \times 10^{-4}$~\cite{PDG2018} and is mainly driven by the Belle~\cite{Kuzmin:2006mw} and \lhcb~\cite{LHCb-PAPER-2014-070} measurements. This value is used as a reference for the measurement of the branching fractions of the decays $B^0_{(s)} \to \Dzb \Kp\Km$. The large contribution from the exclusive decay chain ${\Bz \to D^*(2010)^-\pip}$, ${D^*(2010)^-\to \Dzb \pim}$, with a branching fraction of ${(1.85\pm0.09)\times 10^{-3}}$~\cite{PDG2018}, is not included in the above value. Thus, a  $D^*(2010)^-$ veto is applied. The veto consists of rejecting candidates with $m_{\Dzb\pim}-m_{\Dzb}$  within $\pm4.8$~\mevcc of the expected mass difference~\cite{PDG2018}, which corresponds to $\pm6$ times the \lhcb detector resolution on this quantity. Due to its high production rate and possible misidentification of its decay products, the decay $\Bz \to D^*(2010)^-(\to \Dzb \pim) \pip$ could also contribute as a background to the $B^0_{(s)} \to \Dzb \Kp\Km$ channel. Therefore, the same veto criterion is applied to $B^{0}_{(s)} \to \Dzb \Kp\Km$ candidates as for the $\Bz \to \Dzb \pip \pim$ normalisation mode, where the invariant mass difference $m_{\Dzb\pim}-m_{\Dzb}$ is computed after assigning the pion mass to each kaon in turn.

Only kaon and pion candidates within the kinematic region corresponding to the fiducial acceptance of the RICH detectors~\cite{LHCb-DP-2012-003} are kept for further analysis. This selection is more than $90\%$ efficient for the ${\Bz \to \Dzb \pip \pim}$ signal, as estimated from simulation. Although the $\Dzb$ candidates are selected in a  narrow mass range, studies on simulated samples show a small fraction of ${\Dzb \to \Kp\Km}$ (${\sim 4.5 \times 10^{-5}}$) and ${\Dzb \to \pip\pim}$ (${\sim 3.0 \times 10^{-4}}$) decays, with respect to the genuine $\Dzb \to \Kp\pim$ signal, are still selected. Therefore, loose PID requirements are applied in order to further suppress ${\Dzb \to \Kp\Km}$ and ${\Dzb \to \pip\pim}$ decays. In the doubly Cabibbo-suppressed ${\Dz \to \Kp\pim}$ decay both the kaon and the pion are correctly identified and reconstructed, but the $\Dzb$ flavour is misidentified. This is expected to occur in less than $R_D=(0.348^{+0.004}_{-0.003})\%$~\cite{HFLAV16} of ${\Dzb \to \Kp\pim}$ signal decays. However, such an effect does not impact the measurements of the ratio of branching fractions  ${\BR(\Bz\to\Dzb\Kp\Km)}/{\BR(\Bz\to\Dzb \pip\pim)}$ and ${\BR(\Bs\to\Dzb\Kp\Km)}/{\BR(\Bz\to\Dzb\Kp\Km)}$, as the resulting dilution is the same for the numerator and the denominator.

\subsection{Multivariate selection}
\label{sec:MVA}

Once the initial selections are implemented, a multivariate analysis (MVA) is applied to further discriminate between signal and combinatorial background.  The implementation of the MVA is performed with the TMVA package~\cite{Hocker:2007ht,TMVA4}, using  the $\Bz \to\Dzb \pip\pim$ normalisation channel to optimise the selection. For this purpose only, a loose PID criterion on the pions of the $\pip\pim$ pair is set to reject the kaon and proton hypotheses. The \textit{sPlot} technique~\cite{Pivk:2004ty} is used to statistically separate signal and background in data, with the \Bz candidate invariant mass used as the discriminating variable. The \textit{sPlot weights} (\textit{sWeights})  obtained from this procedure are applied to the candidates to obtain signal and background distributions that are then used to train the discriminant.

To compute the \textit{sWeights}, the signal- and combinatorial-background yields are determined using an unbinned extended maximum-likelihood fit to the invariant-mass distribution of $\Bz$ candidates. The fit uses the sum of a Crystal Ball (CB) function~\cite{CBFunction} and a Gaussian function for the signal distribution and an exponential function for the combinatorial background distribution. The fit is first performed in the invariant-mass range $m_{\Dzb \pip\pim} \in [5240,5420]$~\mevcc, to compute the \textit{sWeights}, and is repeated within the signal region  $[5240,5320]$~\mevcc with all the parameters fixed to the result of the initial fit, except the signal and the background yields, which are found to be $44 \, 690 \pm 540$ and $81 \, 710\pm 570$, respectively. The training samples are produced by applying the necessary signal and background $sWeights$, with half of the data used and randomly chosen for training and the other half for validation.

Several sets of discriminating variables, as well as various linear and non-linear MVA methods, are tested. These variables contain information about the topology and the kinematic properties of the event, vertex quality, $\chi^2_{\rm IP}$ and \pt of the tracks, track multiplicity in cones around the $\Bd$ candidate, relative flight distances between the $\Bd$ and $\Dzb$ vertices and from the PV. All of the discriminating variables have weak correlations ($<1.6\%$) with the invariant mass $m_{\Dzb\pip\pim}$ of the  $\Bz$ candidates. Very similar separation performance is seen for all the tested discriminants. Therefore, a Fisher discriminant~\cite{Fisher:1936et} with the minimal set of the five most discriminating variables is adopted as the default MVA configuration. This option is insensitive to overtraining effects. These five variables are: the smallest values of $\chi^2_{\rm IP}$ and \pt for the tracks of the $\pip\pim$ pair, flight distance significance of the reconstructed $\Bd$ candidates, the $D~\chi^2_{\rm IP}$, and the signed minimum cosine of the angle between the direction  of one of the pions from the $B$ decay and the $\Dzb$ meson, as projected in the plane perpendicular to the beam axis.

\begin{figure}[t]
\centering
\includegraphics[width=12cm]{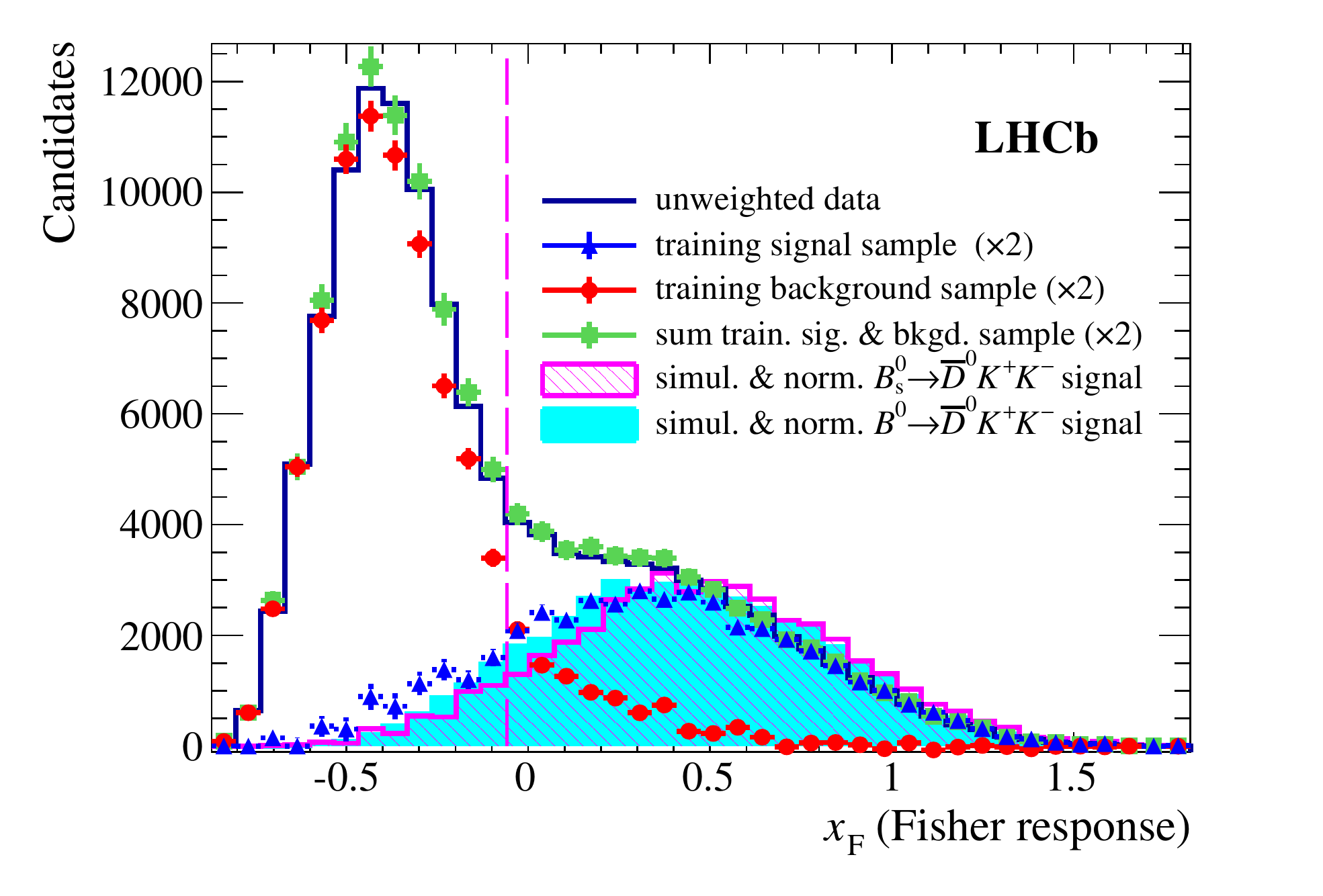}
\caption{Distributions of the Fisher discriminant,  for preselected ${\Bz \to \Dzb \pip \pim}$ data candidates, in the mass range $[5240,5320]$~\mevcc: (black line) unweighted data distribution, and \textit{sWeighted} training samples: (blue triangles) signal, (red circles) background, and (green squares) their sum. The training samples are scaled with a factor of two to match the total yield. The cyan (magenta) filled (hatched) histogram displays the simulated ${\Bz (\Bs) \to \Dzb \Kp \Km}$ decay signal candidates that are normalised to the number of ${\Bz\to\Dzb \pip \pim}$ normalisation channel candidates (blue triangles).  The (magenta) vertical dashed line indicates the position of the nominal selection requirement.}
\label{fig:Fisher}
\end{figure}

Figure~\ref{fig:Fisher} shows the distributions of the Fisher discriminant for the  \textit{sWeighted}  training samples (signal and background) and their sum, compared to the data set of preselected ${\Bz\to\Dzb \pip \pim}$ candidates. These distributions correspond to candidates in the invariant-mass signal region, and agree well within the statistical uncertainties, demonstrating that no overtraining is observed. Based on the fitted numbers of signal  and background candidates, the statistical figure of merit $Q=N_{\rm S}/\sqrt{N_{\rm S}+N_{\rm B}}$ is defined to find an optimal operation point, where $N_{\rm S}$ and $N_{\rm B}$ are the numbers of selected signal and background candidates above a given value ${x_{\rm F}}$ of the Fisher discriminant. The value of ${x_{\rm F}}$ that maximises  $Q$ is found to be $-0.06$, as shown in Fig.~\ref{fig:Fisher} and at this working point the signal efficiency is $(82.4\pm0.4)\%$ and the fraction of rejected background is $(89.2\pm 1.0)\%$.
In Fig.~\ref{fig:Fisher} the distribution  of simulated ${\Bz (\Bs) \to \Dzb \Kp \Km}$ signal decays is also shown to be in good agreement with the \textit{sWeighted} $\Bz \to\Dzb \pip\pim$  data training sample.

\subsection{Particle identification of {\boldmath $h^+h^-$} pairs}
\label{sec:PIDpair}

After the selections, specific PID requirements are set to identify the tracks of the ${B^0_{(s)}}$ decays to distinguish the normalisation channel ${\Bz \to \Dzb \pip\pim}$ and the ${B^0_{(s)}\to \Dzb \Kp\Km}$ signal modes. For the $\Bz \to \Dzb \pip\pim$ normalisation channel, the $\pi^\pm$ candidates must each satisfy  the same PID requirements to identify them as pions, while the kaon and proton hypotheses are rejected. These criteria are tuned by comparing a simulated sample of ${\Bz\to\Dzb \pip\pim}$ signal and a combination of simulated samples that model the misidentified backgrounds. The combination of backgrounds contains all sources expected to give the largest contributions, namely the $\Bz\to\Dzb\Kp\Km$, ${\Bs\to\Dzb\Kp\Km}$, $\Bz\to\Dzb\Kp\pim$, $\Bs\to\Dzb\Km\pip $, $\Lb \to \Dz p\pim$, and $\Lb \to \Dz p\Km$ decays.  The same tuning procedure is repeated for the two ${B^0_{(s)}\to \Dzb \Kp\Km}$ signal modes, where the model for the misidentified background is composed of the main contributing background decays:  ${\Bz\to\Dzb \pip\pim}$, $\Bz\to\Dzb\Kp\pim$, $\Bs\to\Dzb\Km\pip$, $\Lb \to \Dz p\pim$, and $\Lb \to \Dz p\Km$. The $K^\pm$ candidates are required to be positively identified as kaons and the pion and proton hypotheses are excluded. Loose PID requirements are chosen in order to favour the highest signal efficiencies and to  limit possible systematic uncertainties due to data and simulation discrepancies, which arise when computing signal efficiencies related to PID (see Sect.~\ref{sec:Efficiency}).

\subsection{Multiple candidates}
\label{sec:multiCand}

Given the selection described above, $1.2\%$ and $0.8\%$ of the events contain more than one candidate in the $\Bz \to \Dzb\pip\pim$ normalisation and the  $B^0_{(s)}\to\Dzb K^+K^-$ signal modes, respectively. There are  two types of multiple candidates to consider. In the first type, for which two or more good $B$ or $D$ decay vertices are present, the candidate with the smallest sum of the  $B^0_{(s)}$ and \Dzb vertex $\chi^2$ is then kept. In the second type, which  occurs if a swap of the mass hypotheses of the $D$ decay products leads to a good candidate, the PID requirements for the two options  $\Kp\pim$ and $\pip\Km$ are compared and the candidate corresponding to the configuration with the highest PID probability is kept. In order no to bias the $m_{\Dzb h^+ h^-}$ invariant-mass distribution with the choice of the best candidate, it is checked with simulation that the variables used for selection are uncorrelated with the invariant mass, $m_{\Dzb h^+ h^-}$.  It is also computed with simulation that differences between the efficiencies while choosing the best candidate for $\Bz \to \Dzb \pi^+\pi^-$ and $B^0_{(s)}\to\Dzb \Kp \Km$ decays are negligible~\cite{Koppenburg:2017zsh}. 

\section{Fit components and modelling}
\label{sec:BkgModel}

\subsection{Background characterisation}
\label{sec:BkgCharac}
The ${B^0_{(s)}\to \Dzb h^+h^-}$  selected candidates consist of signal and various background contributions:  combinatorial,  misidentified, and partially reconstructed $b$-hadron decays.

The misidentified background originates from real $b$-hadron decays, where at least  one final-state particle is incorrectly identified in the decay chain. For the ${\Bz \to \Dzb \pip\pim}$ normalisation channel, three decays requiring a dedicated modelling are identified: ${\Bz\to\Dzb\Kp\pim}$, ${\Bs\to\Dzb\Km\pip}$, and ${\Lb\to\Dz p\pim}$. Due to the PID requirements, the expected contributions from ${B^0_{(s)} \to \Dzb \Kp\Km}$ are negligible. For the ${B^0_{(s)}\to \Dzb \Kp \Km}$  channels, the modes of interest are ${\Bz\to\Dzb\Kp\pim}$, ${\Bs\to\Dzb\Km\pip}$, ${\Lb\to\Dz p\Km}$, and ${\Lb\to\Dz p\pim}$. Here as well, the contribution from  ${\Bz \to \Dzb \pip \pim}$ is negligible, due to the positive identification of both kaons. Using the simulation and recent measurements for the various branching fractions~\cite{PDG2018,LHCb-PAPER-2013-022,LHCb-PAPER-2013-056,Zupanc:2013iki,LHCb-PAPER-2012-018,LHCb-PAPER-2014-070} and for the fragmentation factors $f_s/f_d$~\cite{fsfd} and $f_{\Lb}/f_d$~\cite{LHCb-PAPER-2014-004}, an estimation of the relative yields with respect to those of the simulated signals is computed over the whole invariant-mass range,  $m_{\Dzb h^+ h^-}\in [5115,6000]$~\mevcc. The values are listed in Table~\ref{tab:Summary_of_Estimates}. The expected yields of the backgrounds related to decays of $\Lb$ baryons cannot be predicted accurately due the limited knowledge of their branching fractions and of the relative production rate $f_{\Lb}/f_d$~\cite{LHCb-PAPER-2014-004}.

\begin{table}[!t]
\centering
\caption{Relative yields, in percent, of the various exclusive $b$-hadron decay backgrounds with respect to that of the ${\Bz\to\Dzb \pip\pi}$ and  ${\B^0_{(s)}\to\Dzb \Kp \Km}$ signal modes. These relative contributions are estimated with simulation in the range $m_{\Dzb h^+ h^-}\in [5115, 6000]$~\mevcc.}
\label{tab:Summary_of_Estimates}
    \begin{tabular}{lcc}
      	\hline
      	\hline \vspace{-0.3cm} \\
      	fraction [$\%$]  & $\Bz \to \Dzb \pip\pim$ & $\B^0_{(s)} \to\Dzb \Kp \Km$   \vspace{0.1cm} \\
      	\hline \vspace{-0.3cm} \\
      $\Bz\to\Dzb\Kp\pim$  & $\al1.3 \pm 0.2$ & $\al2.7 \pm 0.7$ \\
      $\Bs\to\Dzb\Km\pip$ & $\al3.7 \pm 0.7$ & $\al8.1 \pm 2.2$ \\
       $\Lb\to\Dz p\pim$  & $\al3.0 \pm 2.8$ & $\al1.6 \pm 1.7$ \\
      $\Lb\to\Dz p\Km$ & $\al-$ & $\al5.6 \pm 5.4$ \\
      $\Bs\to\Dstarzb \Km\pip$ & $\al1.8\pm0.4$ & $\al8.4 \pm 2.9$\\
      $\Bz\to\Dstarzb [\Dzb\gamma]\pip\pim$ & $16.9\pm2.7$ &  $\al-$\\
      $\Bs\to\Dstarzb [\Dzb\piz] \Kp\Km$ & $\al-$ &$12.8 \pm 6.7$ \\
      $\Bs\to\Dstarzb [\Dzb\gamma] \Kp\Km$ & $\al-$ & $\al5.5 \pm 2.9$ \\
      	\hline
    \end{tabular}
\end{table}

The  partially reconstructed background corresponds to real $b$-hadron decays, where a neutral particle is not reconstructed and possibly one of the other particles is misidentified. For example, ${B^0_{(s)} \to \Dstarzb h^+h^-}$ decays with ${\Dstarzb \to\Dzb\gamma}$ or ${\Dstarzb \to\Dzb \piz}$, where the photon or the neutral pion is not reconstructed. This type of background populates the low-mass region $m_{\Dzb h^+ h^-} < 5240$~\mevcc. For the fit of the ${\Bz \to \Dzb \pip\pim}$ invariant-mass distribution, the main contributions that need special treatment are ${\Bs\to\Dstarzb \Km\pip}$ and ${\Bz\to\Dstarzb [\Dzb\gamma]\pip\pim}$, for which the branching fractions are poorly known~\cite{Satpathy:2002js}. For the  ${B^0_{(s)} \to \Dzb \Kp \Km}$  channels, the decays  ${\Bs\to\Dstarzb \Km\pip}$ and  ${\Bs\to\Dstarzb [\Dzb\pi^0/\gamma] \Kp\Km}$ are of relevance. Using simulation and the available information on the branching fractions~\cite{PDG2018}, and by making the assumption that ${\BR(\Bs\to\Dstarzb \Km\pip)}$ and ${\BR(\Bs\to\Dzb \Km\pip)}$ are equal (this is approximately the case for ${\Bd \to  \Dstarzb\pip\pim}$ and ${\Bd \to \Dzb \pip\pim}$  decays), an estimate of the relative yields with respect to those of the simulated signals is computed over the whole invariant-mass range, ${m_{\Dzb h^+ h^-}\in [5115,6000]}$~\mevcc. The values are given in Table~\ref{tab:Summary_of_Estimates}. The contributions from these backgrounds are somewhat larger than those of the misidentified background, but are mainly located in the mass region ${m_{\Dzb h^+h^-} < 5240}$~\mevcc.

\subsection{Signal modelling}
\label{SignalPDFSection}

The invariant-mass distribution for each of the signal $B^0_{(s)} \to \Dzb h^+h^-$ modes is parametrised with a probability density function (PDF) that is the sum of two CB functions with a common mean,
\begin{equation}
    \mathcal{P}_{\rm sig}(m) = f_{\rm CB}\times {\rm CB}(m; m_0, \sigma_1, \alpha_1, n_1) + (1-f_{\rm CB})\times {\rm CB}(m; m_0, \sigma_2, \alpha_2, n_2).
    \label{2CB_sigPDF}
\end{equation}
The parameters $\alpha_{1,2}$ and $n_{1,2}$  describing the tails of the CB functions are fixed to the values fitted on simulated samples generated uniformly (phase space)
over the  $B^0_{(s)} \to \Dzb h^+h^-$ Dalitz plot. The mean value $m_0$, the resolutions $\sigma_1$ and  $\sigma_2$, and the fraction $f_{\rm CB}$ between the two CB functions are free to vary in the fit to the  ${\Bz \to \Dzb \pip\pim}$  normalisation channel.  For the fit to ${\B^0_{(s)} \to\Dzb\Kp\Km}$ data,  the resolutions $\sigma_1$ and $\sigma_2$  are fixed to those obtained with the normalisation channel, while the mean value $m_0$ and the relative fraction $f_{\rm CB}$ of the two CB  functions are left free. For $\Bs\to\Dzb\Kp\Km$ decays, the same function as for ${\Bd\to\Dzb\Kp\Km}$ is used, the mean values are free but the mass difference between $\Bs$  and  $\Bd$ is fixed to the known value, ${\Delta{m_B} = 87.35 \pm 0.23\mevcc}$~\cite{PDG2018}.

\subsection{Combinatorial background modelling}
\label{CombinatorialPDFSection}

For all channels, the combinatorial background contributes to the full invariant-mass range. It is modelled with an exponential function where the slope $a_{\rm comb.}$ and the normalisation parameter $N_{\rm comb.}$ is free to vary in the fit. The invariant-mass range extends up to 6000\mevcc to include the region dominated by combinatorial background. This helps to constrain the combinatorial background yield and slope.

\subsection{Misidentified and partially reconstructed background modelling}
\label{MisIDPartiallyPDFSection}
The shape of misidentified and partially reconstructed components is modelled by non-parametric PDFs built from large simulation samples. These shapes are determined using the kernel estimation technique~\cite{Cranmer:2000du}. The normalisation of each component is free in the fits. For the normalisation channel $\Bz\to\Dzb \pip\pim$, a component for the decay $\Bz\to\Dstarzb[\Dzb\pi^0]\pip\pim$ is added and modelled by a Gaussian distribution. This PDF also accounts for a possible contribution
from the $\Bp \to \Dzb \pip\pip\pim$ decay, which has a similar shape. In the case of the $B^0_{(s)}\to \Dzb \Kp \Km$  signal channels, the low-mass background also includes a Gaussian distribution  to model the decay $\Bz\to\Dstarzb\Kp\Km$.  To account for differences between data and simulation, these PDFs are modified to match the width and mean of the $m_{\Dzb \pip\pim}$ distribution seen in the data. The normalisation parameter, $N_{{\rm Low}-m}$, of these partially reconstructed backgrounds is free to vary in the fit.

\subsection{Specific treatment of the {\boldmath $\Lb \to \Dz p\pim$}, {\boldmath $\Lb \to \Dz p\Km$}, and {\boldmath $\Xibz\to \Dz p\Km$} backgrounds}
\label{BackgroundWithaProton}

Studies with  simulation show that the distributions of the ${\Lb\to \Dz p\pim}$ and ${\Lb\to\Dz pK^-}$ background modes are broad below the ${B^0_{(s)} \to \Dzb h^+h^-}$ signal peaks. Although their branching fractions have been recently measured~\cite{LHCb-PAPER-2013-056}, the broadness of these backgrounds impacts the determination of both the ${\Bd \to \Dzb h^+h^-}$ and the ${\Bs \to \Dzb h^+h^-}$ signal yields. In particular, knowledge of the ${\Lb\to \Dz p K^-}$  background affects the ${\Bs \to \Dzb \Kp\Km}$ signal yield determination.  The yields of these modes can be determined in data by assigning the proton mass to the ${h^{-}}$ track of the ${B^0_{(s)} \to \Dzb h^+h^-}$ decay, where the charge of ${h^{\pm}}$ is chosen such that it corresponds to the Cabibbo-favoured ${\Dzb}$ mode in the ${\Lb\to\Dz p h^-}$ decay.

The invariant-mass distribution of ${\Lb\to\Dz p\pi^{-}}$ is obtained from the ${\Bz \to \Dzb \pip\pim}$  candidates. A Gaussian distribution is used to model the ${\Lb \to \Dz p\pim}$ signal, while an exponential distribution is used for the combinatorial background. The validity of the background modelling is checked by assigning the proton mass hypothesis to the pion of opposite charge to that expected in the ${\Bz}$ decay. Different fit regions are tested, as well as an alternative fit, where the resolution of the Gaussian PDF that models the ${\Lb \to \Dz p\pim}$ mass distribution is fixed to that of ${\Bz \to \Dzb \pi^+ \pi^-}$. The relative variations of the various configurations are compatible within their uncertainties; the largest deviations are used as the systematic uncertainties.  Finally, the obtained yield for ${\Lb \to \Dz p\pim}$ is ${1101 \pm 144}$,including the previously estimated systematic uncertainties. This yield is then used as a Gaussian constraint in the fit to the ${m_{\Dzb \pip\pim}}$  invariant-mass distribution presented in Sect.~\ref{subsec:NominalFit} and the fit results are presented in Table~\ref{tab:Summary_of_GaussConst}.

The corresponding ${m_{\Dz p \Km}}$ and ${m_{\Dz p \pim}}$ distributions are determined using the ${B^0_{(s)} \to \Dzb \Km\Kp}$ data set. Five components are used to describe the data and to fit the two distributions simultaneously: ${\Lb \to \Dz p\Km}$, ${\Xibz \to \Dz p\Km}$, ${\Lb \to \Dz p \pim}$, ${\Bs \to \Dzb \Km\pip}$, and combinatorial background. A small contribution from the ${\Xibz \to \Dz p\Km}$ decay is observed and is included in the default ${B^0_{(s)} \to \Dzb \Kp \Km}$ fit, where its nonparametric PDF is obtained from  simulation. The ${\Lb \to \Dz p\pim}$ distribution is contaminated by the misidentified backgrounds ${\Lb \to \Dz p\Km}$, ${\Xibz \to \Dz p\Km}$, and ${\Bs \to \Dzb \Km\pip}$ that partially extend outside the fitted region. These yields are corrected according to the expected fractions  as computed from the simulation. The ${\Lb \to \Dz p \Km}$, ${\Xibz \to \Dz p \Km}$, and ${\Lb \to \Dz p\pim}$ signals are modelled with Gaussian distributions, and since the ${\Xibz \to \Dz p\Km}$ yield is small, the mass difference between the ${\Lb}$ and the ${\Xibz}$ baryons is fixed to its known value~\cite{PDG2018}. The effect of the latter constraint is minimal and is  not associated with any systematic uncertainty. The combinatorial background is modelled with an exponential function,  while other misidentified backgrounds are modelled by non-parametric PDFs obtained from  simulation. As for the previous case with ${\Bz \to \Dzb \pip\pim}$  candidates, alternative fits are applied, leading to consistent results where the largest variations are used to assign systematic uncertainties for the determination of the yields of the various components. A test is performed to include a specific cross-feed contribution from the channel ${{\Bs \to \Dzb \Kp\Km}}$. No noticeable effect is observed, except on the yield of the ${\Bs \to \Dzb \Km\pip}$  contribution.  The outcome of this test is  nevertheless included  in the systematic uncertainty.  The obtained yields for the ${\Lb \to \Dz p\Km}$, ${\Xibz\to \Dz p\pim}$, and ${\Lb \to \Dz p\pim}$ decays are ${193 \pm 44}$, ${64 \pm 21}$, and ${74 \pm 32}$ events, respectively, where the systematic uncertainties are included. These yields  and their uncertainties, listed in Table~\ref{tab:Summary_of_GaussConst}, are used as  Gaussian constraints in the fit to the ${{B^0_{(s)} \to \Dzb \Kp\Km}}$ invariant-mass distribution presented in Sect.~\ref{subsec:NominalFit}.

\begin{table}[!t]
\centering
\caption{Fitted yields that are used as Gaussian constraints in the fit to the ${B^0_{(s)} \to \Dzb h^+ h^-}$ invariant-mass distributions presented in Sect.~\ref{subsec:NominalFit}.}
\label{tab:Summary_of_GaussConst}
    \begin{tabular}{ccc}
      	\hline
      	\hline \vspace{-0.3cm} \\
      	Mode  & $\Bz \to \Dzb \pip\pim$ & $\B^0_ {(s)}\to\Dzb \Kp \Km$   \vspace{0.1cm} \\
      	\hline \vspace{-0.3cm} \\
      	 $\Lb \to \Dz p\pim$ & $1101 \pm 144$ & $\al74  \pm 32$ \\
      	$\Lb \to \Dz p\Km$  & $\al-$ & $193  \pm 44$ \\
      	$\Xibz\to \Dz p\pim$ & $\al-$ & $\al64  \pm 21$ \\
      	\hline
    \end{tabular}
\end{table}

\section{Invariant-mass fits and signal yields}
\label{sec:Fit}

\subsection{Likelihood function for the {\boldmath ${B^0_{(s)} \to \Dzb h^+ h^-}$} invariant-mass fit}
\label{subsec:Fonction_likelihood}

The total probability density function  $\mathcal{P}^{\rm tot}_{\theta}(m_{\Dzb h^+ h^-})$ of the fitted parameters $\theta$, is used in the extended likelihood function
\begin{equation}
	\mathcal{L}_{\Dzb h^+ h^-} = \frac{v^n}{n!}e^{-v}\prod_{i=1}^{n} \mathcal{P}^{\rm tot}_\theta(m_{{i},{\Dzb h^+ h^-}}),
	\label{equ:Def_Likelihood}
\end{equation}
where $m_{{i},{\Dzb h^+ h^-}}$ is the invariant mass of candidate $i$, $v$ is the sum of the yields and $n$ the number of candidates observed in the sample.  The likelihood function $\mathcal{L}_{\Dzb h^+ h^-}$ is maximised in the extended fit to the $m_{\Dzb h^+ h^-}$ invariant-mass distribution. The PDF for the $\Bz \to \Dzb\pip\pim$ sample is
\begin{equation}
	\mathcal{P}^{\rm tot}_{\theta}(m_{\Dzb\pip\pim}) = N_{\Dzb\pip\pim} \times \mathcal{P}_{\rm sig}^{\Bd}(m_{\Dzb\pip\pim}) + \sum_{j=1}^{7} N_{j,{\rm bkg}} \times \mathcal{P}_{j,{\rm bkg}}(m_{\Dzb\pip\pim}),
	\label{equ:Def_expression_PDF_tot}
\end{equation}
while that for $B^0_{(s)} \to \Dzb \Kp \Km$ decays is
\begin{eqnarray}
	\mathcal{P}^{\rm tot}_{\theta}(m_{\Dzb\Kp\Km}) &=& N_{\Bz \to \Dzb \Kp \Km} \times \mathcal{P}^{\Bd}_{\rm sig}(m_{\Dzb \Kp\Km})  \\\nonumber
	 &+& N_{\Bs \to \Dzb \Kp \Km} \times \mathcal{P}_{\rm sig}^{\Bs}(m_{\Dzb \Kp\Km}) \\\nonumber
	  &+& \sum_{j=1}^{9} N_{j,{\rm bkg}} \times \mathcal{P}_{j,{\rm bkg}}(m_{\Dzb\Kp\Km}).
	\label{equ:Def_expression_PDF_tot}
\end{eqnarray}
The PDFs used to model the signals $\mathcal{P}_{\rm sig}^{B^0_{(s)}}(m_{\Dzb h^+ h^-})$ are defined by Eq.~\ref{2CB_sigPDF}. The PDFs of each of the seven ($\Bz \to \Dzb\pip\pim$) and nine ($B^0_{(s)} \to \Dzb \Kp\Km$) background components are presented in Sect.~\ref{sec:BkgModel}, while $N_{B^0_{(s)} \to \Dzb h^+ h^-}$ and $N_{j,{\rm bkg}}$ are the signal and background yields, respectively.

\subsection{Default fit and robustness tests}
\label{subsec:NominalFit}

\begin{table}[t]
    \centering
    \caption{Parameters from the default fit to ${\Bz \to \Dzb \pip\pim}$ and ${B^0_{(s)} \to \Dzb \Kp \Km}$ data samples in the invariant-mass range ${m_{\Dzb h^+ h^-}\in[5115, 6000]}$~\mevcc. The quantity $\chi^2/{\rm ndf}$ corresponds to the reduced $\chi^2$ of the fit for the corresponding number of degrees of freedom, ndf, while the  $p$-value is the probability value associated with the fit and is computed with the method of least squares~\cite{PDG2018}.}
    \label{Tab_data_fit_Results}
    \begin{tabular}{ccc}
      		\hline
      		\hline \vspace{-0.3cm} \\
      		Parameter & $\Bz \to \Dzb \pip \pim$ & $B^0_{(s)} \to \Dzb \Kp \Km$  \vspace{0.1cm} \\
      		\hline \vspace{-0.3cm} \\
      		$m_{0}$ [\mevcc]& $5282.0 \pm 0.1$ & $\alm5282.6 \pm 0.3$  \\
      		$\sigma_1$ [\mevcc]& $\al\al\al9.7 \pm 1.0$ & fixed at $9.7$  \\
      		$\sigma_2$ [\mevcc] & $\al\al16.2 \pm 0.8$ & \al fixed at $16.2$ \\
      		$f_{\rm CB}$ & $\al\al\al0.3 \pm 0.1$& $\al\al0.6 \pm 0.1$  \\
      		$a_{\rm comb.}$  [$10^{-3}\times(\mevcc)^{-1}$]&  $\al\all-3.2 \pm 0.1 $  &  $\all-1.3 \pm 0.4$ \\
      \hline \vspace{-0.3cm} \\
      		$N_{\Bz \to\Dzb h^+h^-}$  &  \all$29 \, 943 \pm 243$ &  $1918 \pm 74$ \\
      		$N_{\Bs\to\Dzb h^+h^-}$ &  $\al\al\al-$ &  \al\alll$473 \pm 33$ \vspace{0.3cm} \\
      \hline \vspace{-0.3cm} \\
      		$N_{\rm comb.}$  &  \all$20 \, 266 \pm 463$  &  $\al1720 \pm 231$ \\
      		$N_{\Bs\to \Dzb \Km\pip}$  &  $\al\al\all\all923 \pm 191$  &  $\al151 \pm 47$ \\
      		$N_{\Bd\to \Dzb \Kp\pim}$ &  $\al\all\all2450 \pm 211$  &  $\al131 \pm 65$ \\
      		$N_{\Lb\to \Dz p \Km}$ (constrained) & $\al\al\al-$ &  $\al197 \pm 44$ \\
	     	$N_{\Xibz \to \Dz p\Km}$ (constrained) & $\al\al\al-$ &  $\al\al57 \pm 20$ \\
      		$N_{\Lb\to \Dz p\pim}$ (constrained)&  $\al\all\all1016 \pm 136$   &  $\al\al74 \pm 32$ \\
	        $N_{\Bs\to \Dstarzb \Km\pip}$   & \al\al\all540 (fixed)&  $\al\al833 \pm 185$ \\
      		$N_{\Bs\to \Dstarzb \Kp\Km}$&  $\al\al\al-$    &  $\al\al775 \pm 100$ \\
      		$N_{\Bd\to \Dstarzb [\Dzb\gamma]\pip\pim}$ &  $\al\all\all7697 \pm 325$   &  $\al\al-$ \\
      		$N_{{\rm Low}-m}$ &  \alll$14 \, 914 \pm 222$  &  $1632 \pm 68$ \\
      		\hline \vspace{-0.3cm} \\
      		$\chi^2/{\rm ndf}$ ($p$-value)& 52/46 ($25\%$)  & 43/46 ($60\%$) \\
      		\hline
    \end{tabular}
\end{table}

The default fit to the data is performed, using the MINUIT/MINOS~\cite{James:2004xla} and the RooFit~\cite{Verkerke:2003ir} software packages, in the mass-range $m_{\Dzb h^+ h^-}\in[5115,6000]$~\mevcc.  The fit results are  given in Table~\ref{Tab_data_fit_Results}.

\begin{figure}[t]
    \begin{center}
      \includegraphics[width=12.5cm]{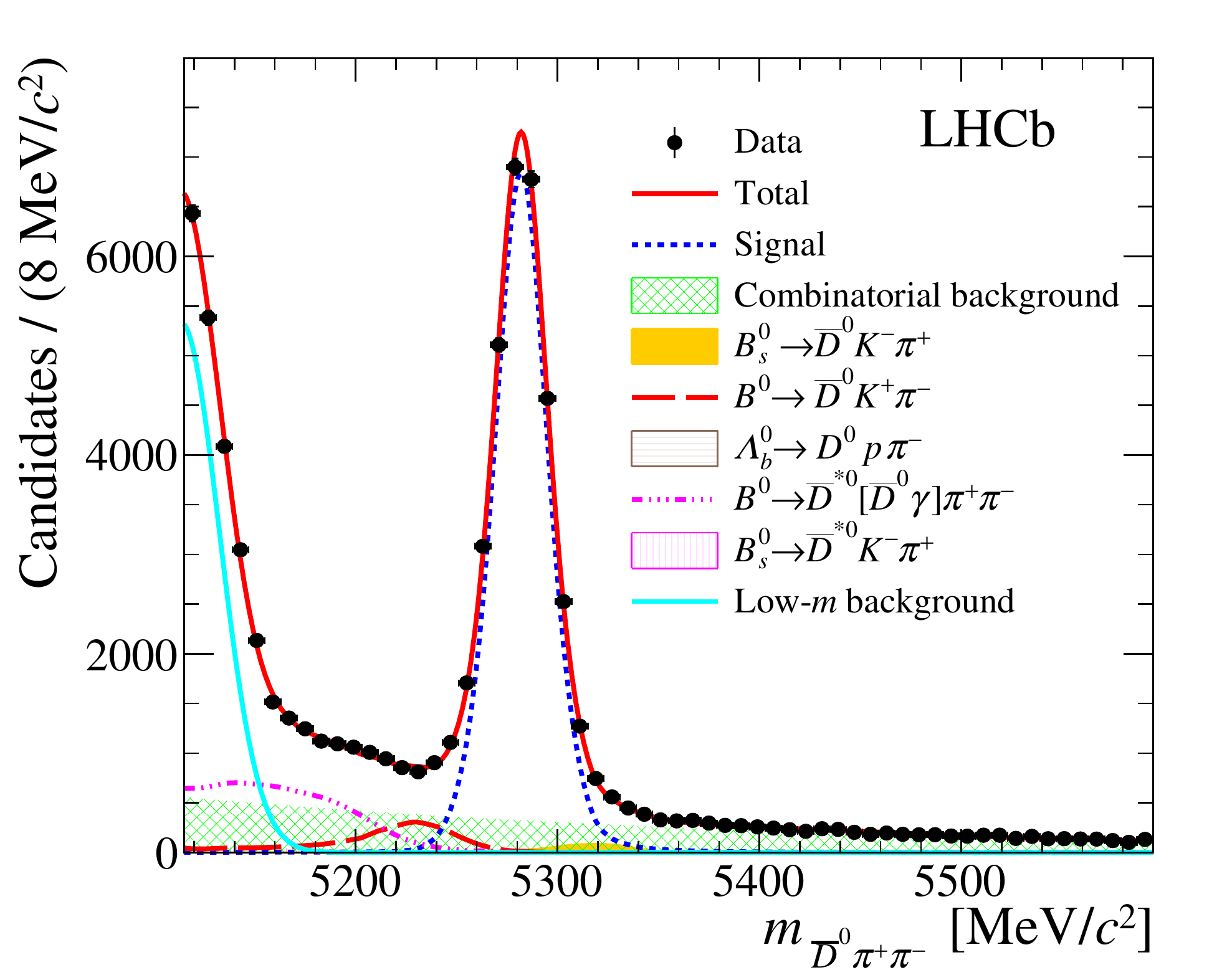}\\
        \includegraphics[width=12.5cm]{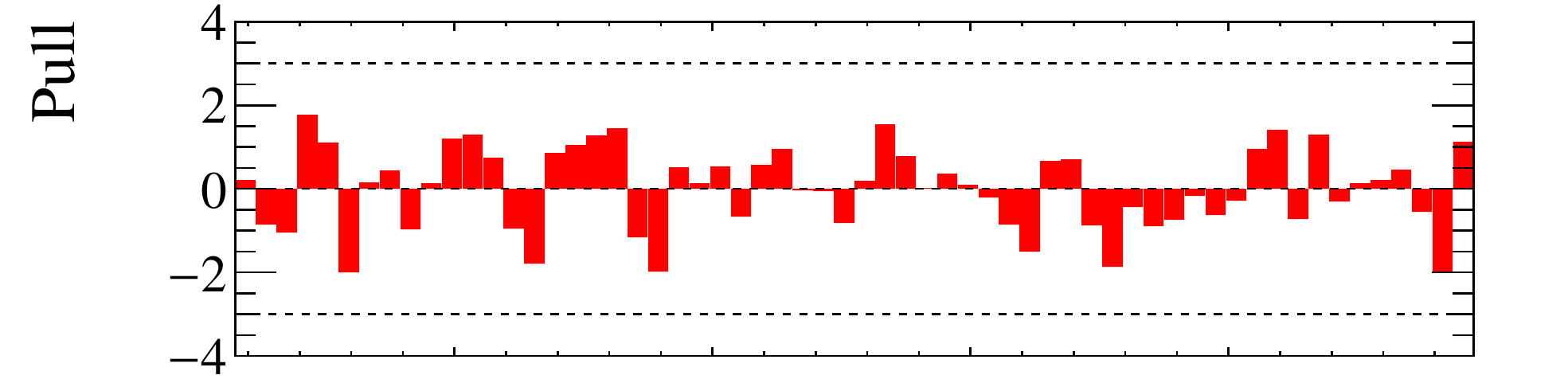}
      \caption{Fit to the  ${m_{\Dzb \pip\pim}}$  invariant-mass distribution with the associated pull plot.}
      \label{Plot_fit_Data_DPiPi}
    \end{center}
\end{figure}

An unconstrained fit to the $m_{\Dzb \pip \pim}$ distribution returns a negative $\Bs\to \Dstarzb \Km\pip$ yield, which is consistent with zero within statistical uncertainties ($-2167\pm 1514$ events), while the expected yield is around $1.8\%$ that of the signal yield, or $540$ events (see Table~\ref{tab:Summary_of_Estimates}). The ${\Bs\to \Dstarzb \Km\pip}$ contribution lies in the lower mass region, where background contributions are complicated, but have little effect on the signal yield determination. In the fit results listed in Table~\ref{Tab_data_fit_Results}, this contribution is fixed to be $540$ events. The difference in the signal yield with and without this constraint amounts to $77$ events, which is included as a systematic uncertainty. The results obtained for the other backgrounds are consistent with the estimated relative yields computed in Sect.~\ref{sec:BkgCharac}. The fit uses Gaussian constraints in the fitted likelihood function for the yields of the modes $\Lb \to \Dz p\Km$, $\Xibz\to \Dz p\pim$, and $\Lb \to \Dz p\pim$, as explained in Sect.~\ref{BackgroundWithaProton}.

\begin{figure}[t]
    \begin{center}
      \includegraphics[width=12.5cm]{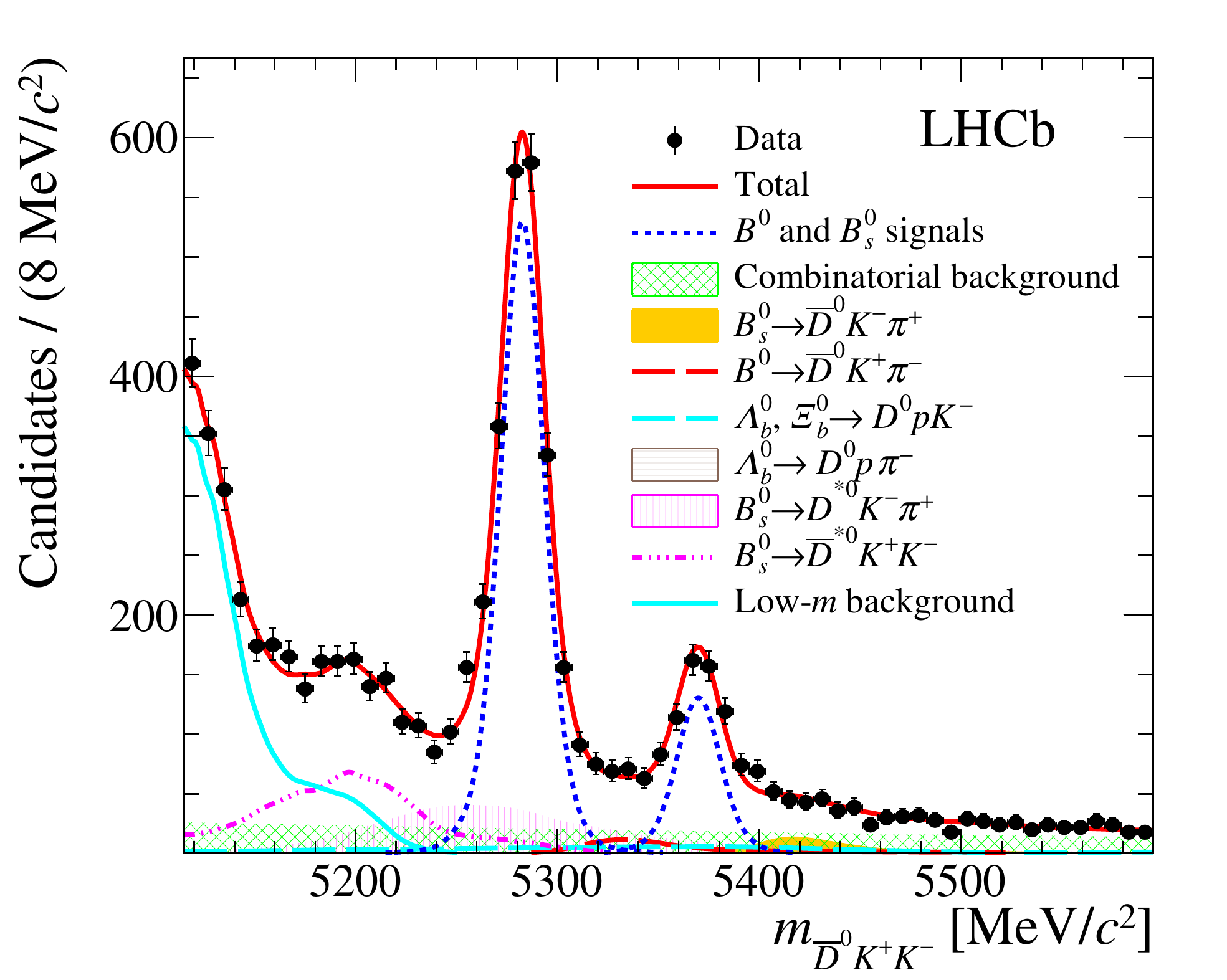}\\
        \includegraphics[width=12.5cm]{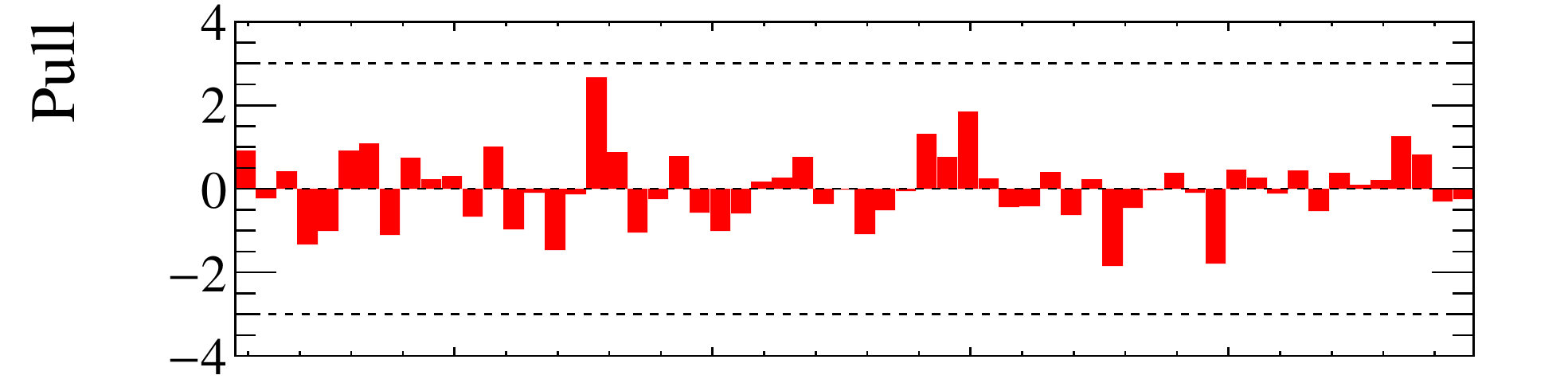}
      \caption{Fit to the  ${m_{\Dzb \Kp\Km}}$  invariant-mass distribution with the associated pull plot.}
      \label{Plot_fit_Data_DKK}
    \end{center}
\end{figure}

\begin{figure}[t]
    \begin{center}

        \includegraphics[width=7.5cm]{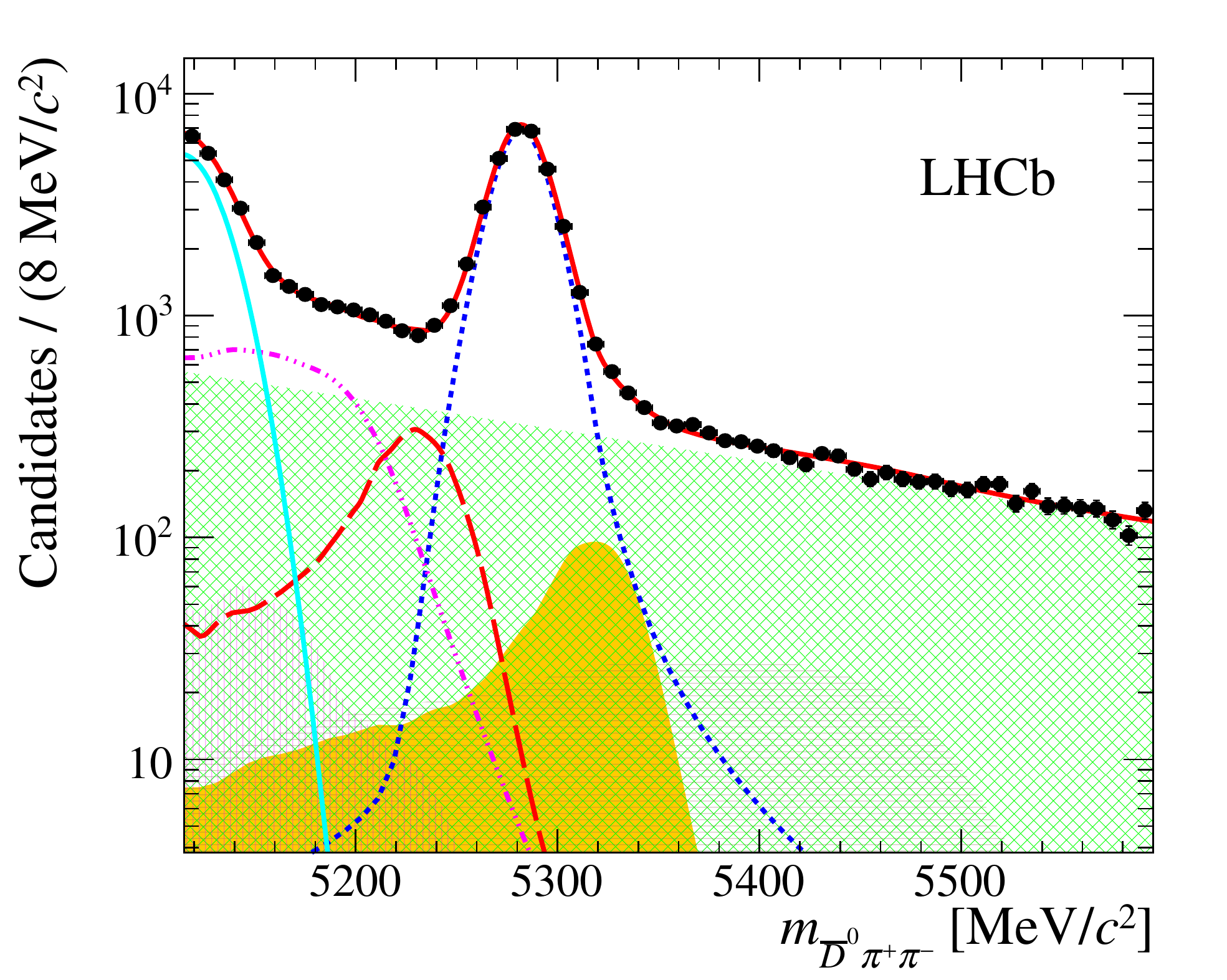}
        \includegraphics[width=7.5cm]{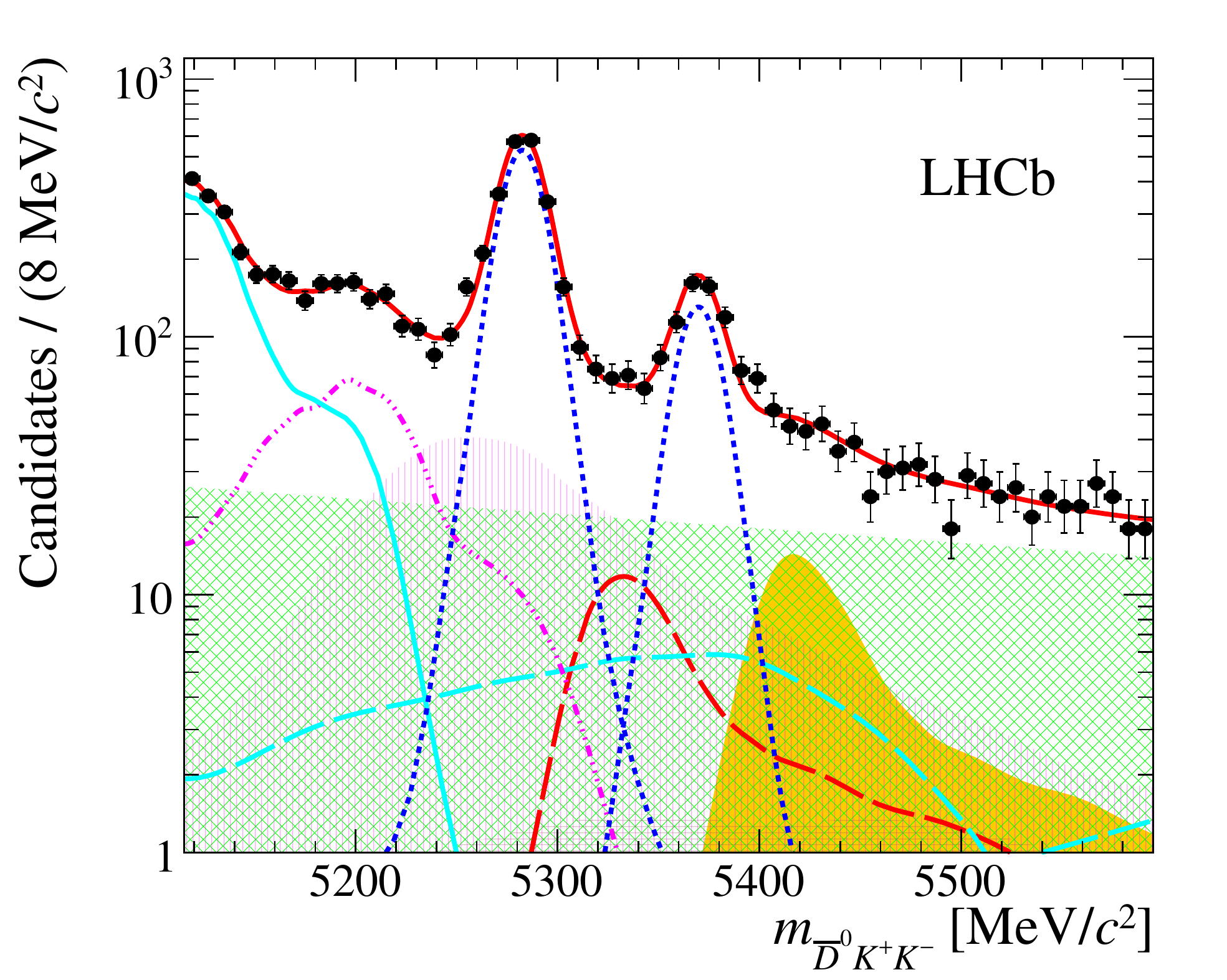}
      \caption{Fit to the (left) ${m_{\Dzb\pip\pim}}$ invariant mass and (right) ${m_{\Dzb \Kp\Km}}$ invariant mass, in logarithmic vertical scale (see the legend on Figs.~\ref{Plot_fit_Data_DPiPi} and~\ref{Plot_fit_Data_DKK}).}
      \label{Plot_fit_Data_Dhh_Log}
    \end{center}
  \end{figure}

The fitted signal yields are ${N_{\Bz\to\Dzb \pip\pim}} =29 \ 943\pm 243$, ${N_{\Bz\to\Dzb\Kp\Km} =1918 \pm 74}$, and ${N_{\Bs\to\Dzb\Kp\Km} = 473 \pm 33}$ events respectively, and the ratio $r_{\Bs/\Bz} \equiv {N_{\Bs\to\Dzb\Kp\Km}}/{N_{\Bz\to\Dzb\Kp\Km}}$ is $(24.7 \pm 1.7)\%$. The ratio $r_{\Bs/\Bz}$ is a parameter in the fit and is used in the computation of the ratio of branching fractions ${{\cal B}\left(\Bs \to \Dzb \Kp\Km\right)}/{{\cal B}\left(\Bd \to \Dzb \Kp\Km\right)}$ (see Eq.~\ref{Ratio_Br_BsoverBd}). The  $\Bs\to\Dzb\Kp\Km$ signal is thus observed with an overwhelming statistical significance. The $\chi^2/{\rm ndf}$ for each fit is very good. The data distributions and fit results are shown in Figs.~\ref{Plot_fit_Data_DPiPi} and  \ref{Plot_fit_Data_DKK}, and Fig.~\ref{Plot_fit_Data_Dhh_Log} shows the same plots with  logarithmic scale in order to visualise the shape and the magnitude of each of the various background components.  The pull distributions, defined as $({n^{\rm fit}_i - n_i })/{\sigma^{\rm fit}_i}$ are also shown in Figs.~\ref{Plot_fit_Data_DPiPi} and  \ref{Plot_fit_Data_DKK}, where the bin number $i$ of the histogram of the $m_{\Dzb h^+h^-}$ invariant mass contains $n_i$ candidates and the fit function yields $n^{\rm fit}_i$  decays, with a statistical uncertainty $\sigma^{\rm fit}_i$. The pull distributions show that the fits are unbiased.

For the ${B^0_{(s)} \to \Dz\Kp\Km}$ channels, the fitted contributions for the ${\Bs\to \Dzb \Km\pim}$ and ${\Bd\to \Dzb \Kp\pip}$ decays are compatible with zero. These components are removed one-by-one in the default fit. The results of these tests are compatible with the output of the default fit. Therefore, no systematic uncertainty is applied.

Pseudoexperiments  are generated using the default fit parameters with their uncertainties (see Table~\ref{Tab_data_fit_Results}), to build 500 (1000) samples of ${\Bz \to \Dzb \pip \pim}$ (${B^0_{(s)} \to \Dzb \Kp \Km}$) candidates according to the yields determined in data. The fit is then repeated on these samples to compute the three most important observables ${N_{\Bz\to\Dzb \pip\pim}}$, ${N_{\Bz\to\Dzb\Kp\Km}}$, and  ${r_{\Bs/\Bz}}$. No bias is seen in  the three considered quantities. A coverage test is performed based on the associated pull distributions yields Gaussian distributions, with the expected mean and standard deviation. This test demonstrates that the statistical uncertainties on the yields obtained from the fit are well estimated.

\section{Calculation of efficiencies and branching fraction ratios}
\label{sec:Efficiency}

The ratios of branching fractions are calculated as
\begin{equation}
 \frac{{\cal B}\left(\Bd \to \Dzb \Kp\Km\right)}{{\cal B}\left(\Bd \to \Dzb \pi^+\pi^-\right)}
= \frac{N_{\Bz\to\Dzb\Kp\Km}}{N_{\Bz\to\Dzb \pip \pim}} \times \frac{\varepsilon_{\Bz\to\Dzb \pip\pim}}{\varepsilon_{\Bz\to\Dzb\Kp\Km}}
\label{Ratio_Br_DKKoverDPiPi}
\end{equation}
and
\begin{equation}
\frac{{\cal B}\left(\Bs \to \Dzb \Kp\Km\right)}{{\cal B}\left(\Bd \to \Dzb \Kp\Km\right)}
=r_{\Bs/\Bz}\times {\frac{\varepsilon_{\Bz\to\Dzb\Kp\Km}}{\varepsilon_{\Bs\to\Dzb\Kp\Km}}} \times {\frac{1}{f_s/f_d}},
\label{Ratio_Br_BsoverBd}
\end{equation}
where the yields are obtained from the fits described in Sect.~\ref{sec:Fit} and  the fragmentation factor ratio $f_s/f_d$ is taken from Ref.~\cite{fsfd}. The efficiencies $\varepsilon$ account for effects related to reconstruction, triggering, PID and selection of the $B^0_{(s)} \to \Dzb h^+h^-$  decays. These efficiencies vary over the Dalitz plot of the  $B$ decays. The total efficiency factorises as
\begin{equation}
\varepsilon_{B^0_{(s)} \to \Dzb h^+h^-}= \varepsilon^{\rm geom}\times\varepsilon^{\rm sel|geom}\times\varepsilon^{\rm PID|sel~\&~geom}\times\varepsilon^{\rm HW~Trig|PID~\&~sel~\&~geom},
      \label{factorisation_effi}
\end{equation}
where ${\varepsilon^{\rm X|Y}}$ is the efficiency of X relative to Y. The contribution $\varepsilon^{\rm geom}$ is determined from the simulation, and corresponds to the fraction of simulated decays which can be fully reconstructed within the \lhcb detector acceptance. The term $\varepsilon^{\rm sel|geom}$ accounts for the software part of the trigger system, the pre-filtering, the initial selection, the Fisher discriminant selection efficiencies, and for the effects related to the reconstruction of the charged tracks. It is computed with simulation, but the part related to the tracking includes corrections obtained from data control samples. The PID selection efficiency $\varepsilon^{\rm PID|sel~\&~geom}$ is determined from the simulation corrected using pure and abundant ${D^*(2010)^+\to\Dz \pip}$ and ${\Lz \to p \pim}$ calibration samples, selected using kinematic criteria only.  Finally, $\varepsilon^{\rm HW~Trig|PID~\&~sel~\&~geom}$ is related to the effects due to the hardware part of the trigger system. Its computation is described in the next section.

As ratios of branching fractions are measured, only the ratios of efficiencies are of interest. Since the multiplicities of all the final states are the same, and the kinematic distributions of the decay products are similar, the uncertainties in the efficiencies largely cancel in the ratios of branching fractions.  The main difference comes from the PID criteria for the $\Bz\to\Dzb \pip \pim$ and  $\Bz\to\Dzb\Kp\Km$ final states.

\subsection{Trigger efficiency}
\label{sec:TrigEfficiency}

The software trigger performance is well described in simulation and is included in $\varepsilon^{\rm sel|geom}$. The efficiency of the hardware trigger depends on data-taking conditions and is determined from calibration data samples. The candidates are of type TOS or TIS, and both types (see Sect.~\ref{sec:Detector}). the efficiency $\varepsilon^{\rm HW~Trig|PID~\&~sel~\&~geom}$ can be written as
\begin{align}
\varepsilon^{\rm HW~Trig|PID~\&~sel~\&~geom} &= \frac{N_{\rm TIS}+N_{\rm TOS\&!TIS}}{N_{\rm ref}} = \varepsilon^{\rm TIS}  + f \times\varepsilon^{\rm TOS}, \label{equa_L0eff}
\end{align}
where $\varepsilon^{\rm TIS} = \frac{N_{\rm TIS}}{N_{\rm ref}}$, $f = \frac{N_{\rm TOS\&!TIS}}{N_{\rm TOS}}$, and $\varepsilon^{\rm TOS}=\frac{N_{\rm TOS}}{N_{\rm ref}}$. The quantity $N_{\rm ref}$ is the number of signal decays that pass all the selection criteria, and $N_{\rm TOS\&!TIS}$ is the number of candidates only triggered by TOS (\ie\ not by TIS). Using Eq.~\ref{equa_L0eff}, the hardware trigger efficiency is calculated from three observables: $\varepsilon^{\rm TIS}$, $f$, and $\varepsilon^{\rm TOS}$.

The quantities  $\varepsilon^{\rm TIS}$ and $f$ are effectively related to the TIS efficiency only. Therefore they are assumed to be the same for the three channels ${B^0_{(s)} \to \Dzb h^+h^-}$ and are obtained from data. The value $f=(69\pm1)\%$ is computed using the number of signal candidates in the ${\Bz \to \Dzb \pip\pim}$ sample obtained from a fit to data for each trigger requirement. The independence of this quantity with respect to the decay channel is checked both in  simulation and in the data with the two  ${B^0_{(s)} \to \Dzb \Kp\Km}$ modes. Similarly, the value of  $\varepsilon^{\rm TIS}$ is found to be $(42.2\pm0.7)\%$.

The efficiency $\varepsilon^{\rm TOS}$ is computed for each of the three decay modes $B^0_{(s)} \to \Dzb h^+h^-$  from phase-space simulated samples
corrected with a calibration data set of  ${\Dstarp\to \Dz[\Km\pip]\pip}$ decays. Studies of the trigger performance~\cite{LHCb:12113055,MartinSanchez:2012yxa} provide a mapping for these corrections as a function of the type of the charged particle (kaon or pion), its electric charge, \pt, the region of the  calorimeter region it impacts, the magnet polarity (up or down), and the time period of data taking (year 2011 or 2012). The value of $\varepsilon^{\rm TOS}$ for each of the three signals is listed in Table~\ref{Tab:Efficiency_Results}.

\begin{table}[t]
\centering
\caption{Total efficiencies ${\varepsilon_{B^0_{(s)} \to \Dzb h^+h^-}}$ and their contributions (before and after accounting for three-body decay kinematic properties) for the each three modes ${\Bz\to\Dzb \pip\pim}$, ${\Bz\to\Dzb \Kp \Km}$, and  ${\Bs\to\Dzb \Kp\Km}$. Uncertainties are statistical only and those smaller than $0.1$ are displayed as $0.1$, but are accounted with their nominal values in the efficiency calculations.}
\label{Tab:Efficiency_Results}
\begin{tabular}{lccc}
\hline
\hline \vspace{-0.3cm} \\
& $\Bz\to\Dzb \pip\pim$ & $\Bz\to\Dzb K^+K^-$ & $\Bs\to\Dzb K^+K^-$  \vspace{0.1cm} \\
\hline \vspace{-0.3cm} \\
$\varepsilon^{\rm geom}$ [$\%$] & $15.8 \pm 0.1$ & $17.0 \pm 0.1$ & $\al16.9 \pm 0.1$ \\
$\varepsilon^{\rm sel~|~geom}$  [$\%$] & $\al1.2 \pm 0.1$ & $\al1.1 \pm 0.1$ & $\al\al1.1 \pm 0.1$ \\
$\varepsilon^{\rm PID~|~sel~\&~geom}$  [$\%$] & $95.5 \pm 1.2$ & $75.7 \pm 1.4$ & $\al76.3 \pm 2.0$ \\
\hline \vspace{-0.3cm} \\
$\varepsilon^{\rm TIS}$ [$\%$] & $42.2\pm0.7$ & $42.2\pm0.7$ &  $\al\all42.2\pm0.7$\\
$\varepsilon^{\rm TOS}$ [$\%$] & $40.6\pm 0.6$ & $40.3 \pm 0.8$ & $\al\all40.6 \pm 1.2$ \\
${\bar{\varepsilon}}^{\rm DP}_{\rm corr.}$ [$\%$] & $85.5 \pm 2.9$  & $95.7 \pm 4.1$ & $101.0^{+3.2}_{-7.1}$ \\
\hline \vspace{-0.3cm} \\
$\varepsilon_{B^0_{(s)} \to \Dzb h^+h^-}^{\rm TIS}$  [$10^{-4}$] & $\al6.4 \pm 0.2$ & $\al5.9 \pm 0.3$ & $\al\al6.0^{+0.3}_{-0.5}$ \\ \vspace{-0.3cm} \\
$\varepsilon_{B^0_{(s)} \to \Dzb h^+h^-}^{\rm TOS}$  [$10^{-4}$] & $\al6.1 \pm 0.2$ & $\al5.7 \pm 0.3 $ & $\al\al5.8 ^{+0.3}_{-0.5}$ \\
\hline \vspace{-0.3cm} \\
$\varepsilon_{B^0_{(s)} \to \Dzb h^+h^-}$ [$10^{-4}$]& $10.6 \pm 0.3$ & $\al9.8 \pm 0.4$ & $\al10.1^{+0.4}_{-0.6}$ \\
\hline
\end{tabular}
\end{table}

\subsection{Total efficiency}
\label{sec:Tot_Efficiency}

The simulated samples used to obtain the total selection  efficiency ${\varepsilon_{B^0_{(s)} \to \Dzb h^+h^-}}$ are generated with phase-space  models for the three-body ${B^0_{(s)} \to \Dzb h^+h^-}$ decays. The three-body distributions in data  are, however, significantly nonuniform (see Sect.~\ref{sec:Results}). Therefore corrections on  ${\varepsilon_{B^0_{(s)} \to \Dzb h^+h^-}}$ are derived to account for the Dalitz plot structures in the considered decays. The relative selection efficiency as a function of the $\Dzb h^+$ and  the $\Dzb h^-$  squared invariant masses, ${\varepsilon(m^2_{\Dzb h^+},m^2_{\Dzb h^-})}$, is determined from simulation and parametrised with a polynomial function of fourth order. The function ${\varepsilon(m^2_{\Dzb h^+},m^2_{\Dzb h^-})}$ is normalised such that its integral is unity over the kinematically allowed phase space. The total efficiency correction ${{\bar{\varepsilon}}^{\rm DP}_{\rm corr.}}$ factor  is calculated, accounting  for the position of each candidate across the Dalitz plot, as
\begin{align}
        {\bar{\varepsilon}}^{\rm DP}_{\rm corr.} & = \frac{\sum_{i} \omega_i}{\sum_{i} {\omega_i/{\varepsilon(m^2_{i,\Dzb h^+},m^2_{i,\Dzb h^-})}}},
\label{Average_eff}
\end{align}
where ${m^2_{i,\Dzb h^+}}$ and ${m^2_{i,\Dzb h^-}}$ are the squared invariant masses of the ${\Dzb h^+}$ and ${\Dzb h^-}$ combinations for the $i^{\rm th}$ candidate in data, and  $\omega_i$ is its signal $sWeight$ obtained from the default fit to the ${B^0_{(s)} \to \Dzb h^+h^-}$  invariant-mass distribution (${m_{B^0_{(s)}}\in[5115,6000]}$~\mevcc).
The statistical uncertainties on the efficiency corrections is evaluated with 1000 pseudoexperiments for each decay mode. The computation of the average efficiency is validated with an alternative procedure in which the phase space is divided into 100 bins for the ${\Bz \to \Dzb\pi^+\pi^-}$ normalisation channel and 20 bins for the ${B^0_{(s)} \to \Dzb \Kp \Km}$ signal modes. This binning is  obtained according to the efficiency map of each decay, where areas with similar efficiencies are grouped together. The total average efficiency is then computed as a function of the efficiency and the number of candidates in each bin. The two methods give compatible results within the uncertainties. The values of ${{\bar{\varepsilon}}^{\rm DP}_{\rm corr.}}$ for each of the three signals are listed in Table~\ref{Tab:Efficiency_Results}.

Table~\ref{Tab:Efficiency_Results} shows the value of the total efficiency ${\varepsilon_{B^0_{(s)} \to \Dzb h^+h^-}}$ and its contributions. The relative values of ${\varepsilon_{B^0_{(s)} \to \Dzb h^+h^-}^{\rm TIS}}$ and ${\varepsilon_{B^0_{(s)} \to \Dzb h^+h^-}^{\rm TOS}}$, for TIS and TOS triggered candidates, are also given. The total efficiency is obtained as (see Eq.~\ref{equa_L0eff})
\begin{align}
{\varepsilon_{B^0_{(s)} \to \Dzb h^+h^-}}= {\varepsilon_{B^0_{(s)} \to \Dzb h^+h^-}^{\rm TIS}} +f \times {\varepsilon_{B^0_{(s)} \to \Dzb h^+h^-}^{\rm TOS}},
\end{align}
where $f=(69\pm1)\%$. The total efficiencies for the three ${B^0_{(s)}\to \Dzb h^+h^-}$  modes are compatible within their uncertainties.

\section{Systematic uncertainties}
\label{sec:Systematic}

Many sources of systematic uncertainty cancel in the ratios of branching fractions. Other sources are described below.

\subsection{Trigger}
The calculation of the hardware trigger efficiency is described in Sect.~\ref{sec:TrigEfficiency}. To determine ${\varepsilon^{\rm HW~Trig|PID~\&~sel~\&~geom}}$, a data-driven method is exploited. It is based on $\varepsilon^{\rm TOS}$, as described in~Refs.\cite{MartinSanchez:2012yxa} and~\cite{LHCb-PAPER-2014-028}, and on the quantities $f$ and $\varepsilon^{\rm TIS}$, determined on the data normalisation channel ${\Bd\to\Dzb \pip\pim}$  (see Eq.~\ref{equa_L0eff}). The latter two quantities depend on the TIS efficiency of the hardware trigger and are assumed to be the same for all three modes. The values of $f$ and $\varepsilon^{\rm TIS}$ are consistent for the ${\Bd\to\Dzb\Kp\Km}$ and the ${\Bs\to\Dzb\Kp\Km}$  channels; no systematic uncertainty is assigned for this assumption. Simulation studies show that these values are consistent for ${\Bd\to\Dzb\Kp\Km}$ and ${\Bd\to\Dzb \pip \pim}$ channels. A  $2.0\%$ systematic uncertainty, corresponding to the maximum observed deviation with  simulation, is assigned on the ratio of their relative ${\varepsilon^{\rm HW~Trig|PID~\&~sel~\&~geom}}$  efficiencies.

\subsection{PID}

A systematic uncertainty is associated to the efficiency $\varepsilon^{\rm PID|sel~\&~geom}$ when final states of the signal and normalisation channels are different. For each track which differs in the signal channel ${\Bz\to\Dzb \Kp \Km}$ and the normalisation channel $\Bz\to\Dzb \pip \pim$, an uncertainty of 0.5\% per track due to the kaon or pion  identification requirement is applied (\eg see Refs.~\cite{LHCb-PAPER-2011-022,LHCb-PAPER-2013-022}). As the same PID requirements are used for \Dzb decay products for all modes, the charged tracks from those decay products do not need to be considered. The relevant systematic uncertainties are added linearly to account for correlations in these uncertainties. An overall PID systematic uncertainty of $2.0\%$  on the ratio ${\BR(\Bz\to\Dzb \Kp \Km)}/{\BR(\Bz\to\Dzb \pip \pim)}$ is assigned.

\subsection{Signal and background modelling}

Systematic effects due to the imperfect modelling of both the signal and background distributions in the fit to $m_{\Dzb h^+ h^-}$ are studied. Additional components are considered for each fit on $m_{\Dzb \pip\pim}$ and $m_{\Dzb \Kp \Km}$. Moreover the impact of backgrounds with a negative yield, or compatible with zero at one standard deviation is evaluated. The various sources of systematic uncertainties discussed in this section are given in Table~\ref{tab:Recap_Syst_Dhh_fit}. The main sources are related to resolution effects and to the modelling of the signal and background PDFs.
	
A systematic uncertainty is assigned for the modelling of the PDF $\mathcal{P}_{\rm sig}$, defined in Eq.~\ref{2CB_sigPDF}. The value of the tail parameters $\alpha_{1,2}$ and  $n_{1,2}$  are fixed to those obtained from simulation. To test the validity of this constraint, new sets of tail parameters, compatible with the covariance matrix obtained from a fit to simulated signal decays, are generated and used as new fixed values. The variance of the new fitted yields is $1.0\%$ of the yield ${N_{\Bz\to\Dzb\pip\pim}}$, which is taken as the associated systematic uncertainty.
For the fit to the ${B^0_{(s)} \to \Dzb \Kp\Km}$ candidates, the above changes to the tail parameters correspond to a $1.4\%$ relative effect on the yield ${N_{\Bz\to\Dzb\Kp\Km}}$ and $0.4\%$ on the ratio $r_{\Bs/\Bz}$.  Another systematic uncertainty is linked to the relative resolution of the ${\Bs\to\Dzb \Kp \Km}$ mass peak with respect to that of the ${\Bz\to\Dzb \Kp \Km}$ signal. In the default fit, the resolutions of these two modes are fixed to be the same. Alternatively, the relative difference of the resolution for the two modes can be taken to be proportional to the  kinetic energy released in the decay, ${Q_{d,(s)} = m_{B^0_{(s)}} - m_{\Dzb} - 2m_{X}}$, where $m_X$ indicates the known mass of the $X$ meson, so that the resolution of the \Bz signal stays unchanged, while that of the \Bs distribution is multiplied by ${Q_s / Q_d = 1.02}$. The latter effect results in a small change of $0.2\%$ on ${N_{\Bz\to\Dzb\Kp\Km}}$, as expected, and a larger variation of $1.7\%$ on  $r_{\Bs/\Bz}$. A third systematic uncertainty on $B^0_{(s)} \to \Dzb \Kp \Km$ signal modelling is computed to account for the mass difference ${\Delta m_B}$ which is fixed in this fit (see Sect.~\ref{SignalPDFSection}). When left free in the fit, the measured mass difference ${\Delta m_B = 88.29\pm 1.23\mevcc}$ is consistent with the value fixed in the default fit, which creates a relative change of $1.6\%$ on $N_{\Bz\to\Dzb\Kp\Km}$ and a larger one of $3.8\%$ on $r_{\Bs/\Bz}$. These three sources of systematic uncertainty on the ${B^0_{(s)} \to \Dzb \Kp \Km}$ invariant-mass modelling are considered as uncorrelated, and are added in quadrature to obtain a global relative systematic uncertainty of $2.1\%$ on the yield ${N_{\Bz\to\Dzb\Kp\Km}}$ and $4.2\%$ on the ratio $r_{\Bs/\Bz}$.

\begin{table}[t]
    \centering
\caption{Relative systematic uncertainties, in percent, on ${N_{\Bz\to\Dzb \pip\pim}}$, ${N_{\Bz\to\Dzb \Kp \Km}}$ and the ratio ${N_{\Bz\to\Dzb \pip\pim}}$/${N_{\Bz\to\Dzb \Kp \Km}}$ and ${r_{\Bs/\Bz}}$, due to PDFs modelling in the $m_{\Dzb\pip\pim}$ and $m_{\Dzb \Kp \Km}$ fits. The uncertainties are uncorrelated and summed in quadrature.}
\label{tab:Recap_Syst_Dhh_fit}
    \begin{tabular}{cccc}
      	\hline
      	\hline \vspace{-0.3cm} \\
      	Source & $N_{\Bz\to\Dzb \pip\pim}$ & $N_{\Bz\to\Dzb \Kp \Km}$ & $r_{\Bs/\Bz}$ \vspace{0.1cm}  \\
      	\hline \vspace{-0.3cm} \\
      	${B^0_{(s)} \to\Dzb h^+h^-}$  signal PDF & 1.0 & 2.1 & 4.2 \\
      	\hline \vspace{-0.3cm} \\
      	${\Bz\to\Dstarzb [\Dzb\gamma] \pip\pim}$  & 1.6 & $-$& $-$\\
      	${\Bz\to\Dzb\Kp\pim}$  & 0.3 & $-$& $-$\\
      	${\Bs\to\Dstarzb\Km\pip}$  & 0.4 & 1.4 & 0.4 \\
      	${\Bs\to\Dstarzb\Kp\Km}$  & $-$& 0.5 & 1.3 \\
      	\hline \vspace{-0.3cm} \\
      	Smearing \& shifting  & 0.5 & 0.1 & 0.9 \\
      	\hline \vspace{-0.3cm} \\
      	Total & 2.0 & 2.6 & 4.5 \\
      	\hline \vspace{-0.3cm} \\
      	Total on $N_{\rm sig} / N_{\rm normal}$ & \multicolumn{2}{c}{3.2} & 4.5 \\
      	\hline
    \end{tabular}
\end{table}

For the default fit on $m_{\Dzb\pip\pim}$ (see Table~\ref{Tab_data_fit_Results}), the ${\Bz\to\Dstarzb [\Dzb\gamma] \pip\pim}$ and ${\Bz\to\Dzb\Kp\pim}$ components are the main peaking backgrounds and the contribution from ${\Bs\to\Dstarzb\Km\pip}$ is fixed to  the expected value from simulation. The  ${\Bz\to\Dstarzb [\Dzb\gamma] \pip\pim}$ background is modelled in the default fit with a nonparametric PDF determined on a phase-space simulated sample of ${\Bz\to\Dstarzb [\Dzb\gamma] \pip\pim}$ decays. In an alternative approach, the modelling of that background  is replaced by nonparametric PDFs determined from  simulated samples of  ${\Bz\to\Dstarzb [\Dzb\gamma] \rho^0}$ decays  with various polarisations. Two values for the longitudinal polarisation fraction are tried, one from the  colour-suppressed mode ${\Bz\to \Dstarzb \omega}$, $f_{\rm L} = (66.5 \pm 4.7 \pm 1.5)\%$~\cite{Lees:2011gw} (this result is consistent with the result presented in Ref.~\cite{Matvienko:2015gqa}) and the other from the colour-allowed mode ${B^0\to \Dstarm \rhop}$, $f_{\rm L} = (88.5 \pm 1.6 \pm 1.2)\%$~\cite{Csorna:2003bw}. A systematic uncertainty of $1.6\%$ for the ${\Bz\to\Dstarzb [\Dzb\gamma] \pip\pim}$ modelling, corresponding to the largest deviation from the nominal result, is assigned. A different model of simulation for the generation of the background ${\Bz\to\Dzb\Kp\pim}$ decays is used to define the nonparametric PDF used in the invariant-mass fit. The first is a phase-space model where the generated signals decays are uniformly distributed over a regular-Dalitz plot, while the other is uniformly distributed over the square version of the Dalitz plot. The definition of the square-Dalitz plots is given in Ref.~\cite{LHCb-PAPER-2014-070}. The difference between these two PDFs for the ${\Bz\to\Dzb\Kp\pim}$ background corresponds to a $0.3\%$ relative effect. The component ${\Bs\to\Dstarzb\Km\pip}$ is found to be  initially negative (and compatible with zero) and then fixed in the default fit, resulting  in a relative systematic uncertainty of $0.4\%$.

The main background channels in the fit to $m_{\Dzb \Kp \Km}$ are ${\Bs\to\Dstarzb \Kp \Km}$ and ${\Bs\to\Dstarzb \Km\pip}$. The nonparametric PDF for ${\Bs\to\Dstarzb\Kp\Km}$ decays is computed from an alternative simulated sample, where the nominal phase-space simulation is replaced by that computed with a square-Dalitz plot generation of the simulated decays. The measured difference between the two models results in relative systematic uncertainties on $N_{\Bz\to\Dzb\Kp\Km}$ and $r_{\Bs/\Bz}$ of $0.5\%$ and $1.3\%$, respectively. The component ${\Bs\to\Dstarzb \Km\pip}$ is modelled with a nonparametric PDF from the square-Dalitz plot simulation. Alternatively, the PDF of the  ${\Bs\to\Dstarzb \Km\pip}$ background is modelled with a nonparametric PDF determined from a simulated sample of ${\Bs\to\Dstarzb \Kstarzb}$ decays, with polarisation taken from the similar mode ${\Bp \to \Dstarzb \Kstarp}$, $f_{\rm L} = (86 \pm 6 \pm 3)\%$~\cite{Aubert:2003ae}. The difference obtained for these two PDF models for the ${\Bs\to\Dstarzb \Km\pip}$ background  gives  relative systematic uncertainties on $N_{\Bz\to\Dzb\Kp\Km}$ and $r_{B_s/B_d}$ equal to $1.4\%$ and $0.4\%$.

Systematic uncertainties for the constrained $\Lb \to \Dz p\Km$ or $\Lb \to \Dz p\pim$  and  $\Xibz\to \Dz p\pim$ decay  yields are discussed in Sect.~\ref{BackgroundWithaProton} and are already taken into account when fitting the ${B^0_{(s)}\to \Dzb h^+h^-}$ invariant-mass distributions.

Finally, the impact of the simulation tuning that is described in Sect.~\ref{MisIDPartiallyPDFSection} is evaluated by performing the default fit  without modifying the PDFs of the various backgrounds to match the width and mean invariant masses seen in data. The resulting discrepancies give a relative effect of $0.5\%$ on ${N(\Bz\to\Dzb \pip \pim)}$, $0.1\%$ on ${N(\Bz\to\Dzb \Kp \Km)}$, and $0.9\%$ on $r_{\Bs/\Bz}$.

\begin{table}[b]
\centering
\caption{Relative systematic uncertainties, in percent,  on the ratio of branching fractions $\mathcal{R}_{\Dzb \Kp\Km/\Dzb \pip\pim}$ and $\mathcal{R}_{\Bs/\Bz}$. The uncertainties are uncorrelated and summed in quadrature.}
\label{tab:Systematics_Ratios_Summary}
\begin{tabular}{ccc}
    \hline
    \hline \vspace{-0.3cm} \\
    Source & $\mathcal{R}_{\Dzb \Kp\Km/\Dzb \pip\pim}$ & $\mathcal{R}_{\Bs/\Bz}$ \vspace{0.1cm} \\
    \hline \vspace{-0.3cm} \\
    HW trigger efficiency & 2.0  & $-$\\
    PID efficiency  & 2.0 & $-$ \\
    PDF modelling & 3.2 & 4.5\\
    $f_s/f_d$ & $-$ & 5.8 \\
    \hline \vspace{-0.3cm} \\
    Total & 4.3 & 7.3 \\
    \hline
\end{tabular}
\end{table}

\subsection{Summary of systematic uncertainties}
\label{Sec::Syst_summary}

The systematic uncertainties contributing to the ratio of branching fractions ${\mathcal{R}_{\Dzb \Kp\Km/\Dzb \pip\pim} \equiv {\BR(\Bz\to \Dzb \Kp\Km)}/{\BR(\Bz\to\Dzb \pip\pim)}}$ (see Eq.~\ref{Ratio_Br_DKKoverDPiPi}) and for the ratio ${\mathcal{R}_{\Bs/\Bz} \equiv {\BR(\Bs\to \Dzb \Kp\Km)}/{\BR(\Bz\to\Dzb\Kp\Km)}}$ (see Eq.~\ref{Ratio_Br_BsoverBd}) are listed in Table~\ref{tab:Systematics_Ratios_Summary}. All sources of systematic uncertainties are uncorrelated and are therefore summed in quadrature. For the  ratio $\mathcal{R}_{\Bs/\Bz}$  the external input ${f_s/f_d = 0.259 \pm 0.015}$~\cite{fsfd} introduces the dominant systematic uncertainty of $5.8\%$.

\section{Results}
\label{sec:Results}

The ratios of branching fractions are measured to be
\begin{equation}
	\frac{\BR(\Bz\to\Dzb\Kp\Km)}{\BR(\Bz\to\Dzb \pip\pim)} = (6.9 \pm 0.4 \pm 0.3)\%
	\label{equ:Conclusion_Rap_B2DKKoverB2DPiPi}
\end{equation}
and
\begin{equation}
    \frac{\BR(\Bs\to\Dzb\Kp\Km)}{\BR(\Bz\to\Dzb\Kp\Km)} = (93.0 \pm 8.9 \pm 6.9)\%,
	\label{equ:Conclusion_Rap_Bs2DKKoverB2DK^+ K-}
\end{equation}
where the first uncertainties are statistical and the second are systematic. Using the branching fraction ${\BR(\Bz\to\Dzb \pip\pim) = (8.8 \pm 0.5) \times 10^{-4}}$~\cite{PDG2018}, the branching fraction of the ${\Bz\to\Dzb\Kp\Km}$ decay is measured to be
\begin{equation}
		\BR(\Bz\to\Dzb\Kp\Km) = (6.1 \pm 0.4 \pm 0.3 \pm 0.3) \times 10^{-5},
		\label{equ:Conclusion_BR_B2DK^+ K-}
\end{equation}
where the third uncertainty is due to the limited knowledge of  $\BR(\Bz\to\Dzb \pip\pim)$. The  branching ratio of the decay ${\Bs\to\Dzb\Kp\Km}$ is measured to be
\begin{equation}
		\BR(\Bs\to\Dzb\Kp\Km) = (5.7 \pm 0.5 \pm 0.4 \pm 0.5) \times 10^{-5},
		\label{equ:Conclusion_BR_Bs2DK^+ K-}
\end{equation}
where the third uncertainty is due to the limited knowledge of $\BR(\Bz\to\Dzb \Kp \Km)$.
These results are compatible with and more precise than the previous \lhcb results~\cite{LHCb-PAPER-2012-018} for the same decays, \ie ${\cal B}\left(\Bd \to \Dzb \Kp\Km\right) =$ $ (4.7 \pm 0.9 \pm 0.6 \pm 0.5)\times 10^{-5}$ and ${\cal B}\left(\Bs \to \Dzb \Kp\Km\right)=$ $(4.2 \pm 1.3 \pm 0.9 \pm 1.1)\times 10^{-5}$, which were based on a subset of the current data set. The measurement of the branching ratios $\BR({B^0_{(s)} \to\Dzb\Kp\Km})$ is the first step towards a Dalitz plot analysis of these modes using the LHC Run-2 data sample. Nonetheless, an inspection of the Dalitz plot is performed and several structures are visible in the  ${\Bz\to\Dzb \Kp \Km}$ and  ${\Bs\to\Dzb \Kp \Km}$ decays.

\begin{figure}[t]
\begin{center}
    \includegraphics[width=10cm]{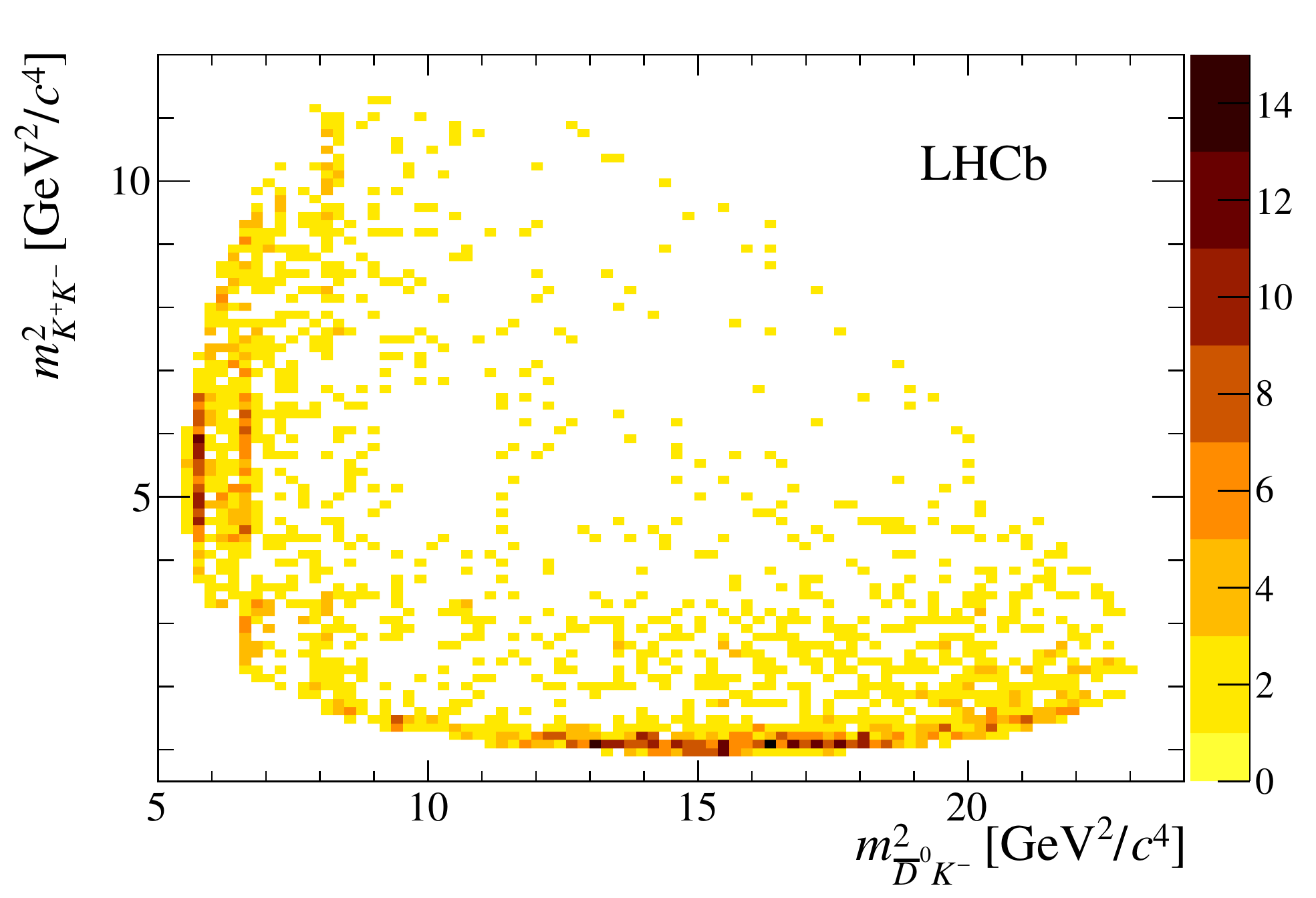}
    \caption{Dalitz plot for ${\Bz\to\Dzb\Kp\Km}$ candidates in the signal region ${m_{\Dzb\Kp \Km}\in [5240,5320]\mevcc}$.}
    \label{fig:DP_DKvsKK_B_region_BtoDKK_5juin2015}
\end{center}
\end{figure}

\begin{figure}[b]
\begin{center}
    \includegraphics[width=10cm]{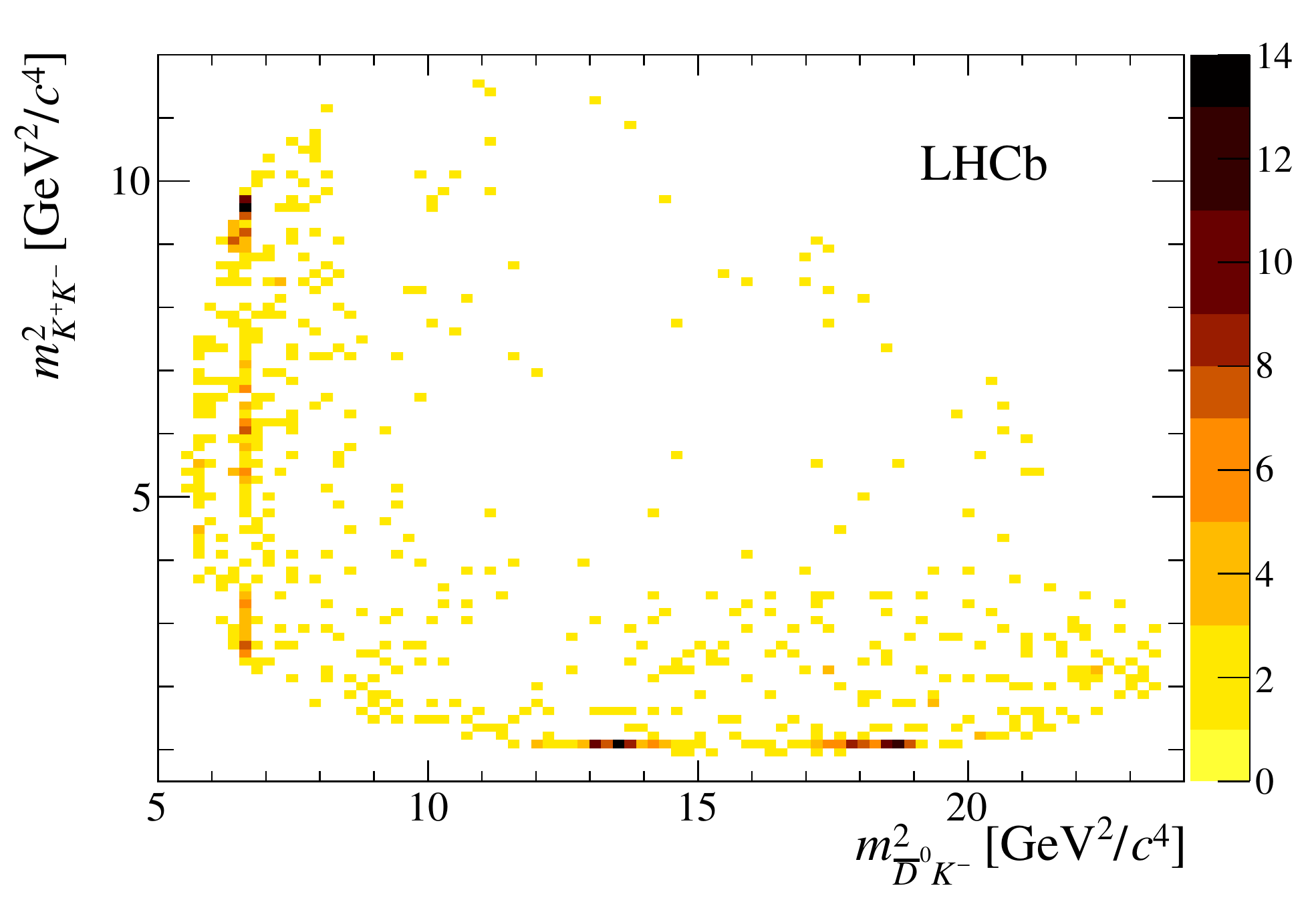}
    \caption{Dalitz plot for ${\Bs\to\Dzb\Kp\Km}$ candidates in the signal region ${m_{\Dzb\Kp \Km}\in [5327,5407]\mevcc}$.}
    \label{fig:DP_DKvsKK_Bs_region_BstoDKK_5juin2015}
\end{center}
\end{figure}

The Dalitz plot $(m^2_{\Dzb \Km},m^2_{\Km \Kp})$ distribution of $\Bd \to \Dzb \Kp\Km$ candidates populating the \Bz signal mass range, $m_{\Dzb\Kp \Km}\in [5240,5320]$~\mevcc (\ie $\pm40$~\mevcc around the $\Bz$ mass), is displayed in Fig.~\ref{fig:DP_DKvsKK_B_region_BtoDKK_5juin2015}. Several resonances are clearly visible. In the $\Kp\Km$ system, some unknown combination of the resonances ${a_0}(980)$ and $f_0(980)$ seem to be dominant. The search for the rare $\Bz\to\Dzb\phi$  decay using the same data sample is described in a separate publication~\cite{LHCb-PAPER-2018-015}. For the $\Dzb \Km$ system, the first band below $6 \ {\mathrm {GeV}}^2/c^4$ corresponds to the partially reconstructed decay ${\Bs \to D_{s1}(2536)^- \Kp/\pip}$, with ${D_{s1}(2536)^-\to\Dstarzb\Km}$ (\ie\ a background component due to the decay $\Bs \to \Dstarzb \Km\Kp$ or $\Bs \to \Dstarzb \Km\pip$, with the pion misidentified). The decay ${D_{s1}(2536)^- \to\Dzb\Km}$ is forbidden by the conservation of parity in strong interactions and cannot explain the observed feature. The second band around $6.6 \ {\mathrm {GeV}}^2/c^4$  is related to the mode ${\Bz \to D^*_{s2}(2573)^- \Kp}$, with ${D^*_{s2}(2573)^-\to\Dzb\Km}$ and a third vertical band can be distinguished at about $8.2 \ {\mathrm {GeV}}^2/c^4$ which corresponds to a potential superposition of the  $D_{s1}^*(2860)^-$ and the  $D_{s3}^*(2860)^-$ resonances previously observed by \lhcb~\cite{LHCb-PAPER-2014-035,LHCb-PAPER-2014-036}.

The Dalitz plot $(m^2_{\Dzb \Km},m^2_{\Km \Kp})$  distribution of $B^0_{(s)} \to  \Dzb \Kp\Km$ candidates populating the \Bs signal mass range, ${m_{\Dzb \Kp \Km}\in [5327,5407]\mevcc}$ (\ie $\pm40$~\mevcc around the $\Bs$ mass), is shown in Fig.~\ref{fig:DP_DKvsKK_Bs_region_BstoDKK_5juin2015}. Again, several resonances can be clearly identified. In the $\Kp\Km$ system, the  $\phi$ resonance is observed and the study of the corresponding decay is presented in a separate publication~\cite{LHCb-PAPER-2018-015}. There is some possible accumulation of candidates in a broad structure around $1.7$~\gevcc, which may correspond to the $\phi(1680)$ state. In addition, in the $\Dzb\Km$ system, the $D^*_{s2}(2573)^-$ resonance is identifiable.

An analysis with additional \lhcb data will enable the study of $D_s^{**}$ spectroscopy, particularly those resonances that are natural spin-parity members of the $1D$ and $1F$ families. The differences between the $\Bz$ and $\Bs$ modes are also interesting. In addition, different resonances can contribute strongly with respect to $\Bs\to \Dzb \Km\pip$ decays~\cite{LHCb-PAPER-2014-035,LHCb-PAPER-2014-036}.

\section*{Acknowledgements}
%
%
\noindent We express our gratitude to our colleagues in the CERN
accelerator departments for the excellent performance of the LHC. We
thank the technical and administrative staff at the LHCb
institutes.
We acknowledge support from CERN and from the national agencies:
CAPES, CNPq, FAPERJ and FINEP (Brazil); 
MOST and NSFC (China); 
CNRS/IN2P3 (France); 
BMBF, DFG and MPG (Germany); 
INFN (Italy); 
NWO (Netherlands); 
MNiSW and NCN (Poland); 
MEN/IFA (Romania); 
MinES and FASO (Russia); 
MinECo (Spain); 
SNSF and SER (Switzerland); 
NASU (Ukraine); 
STFC (United Kingdom); 
NSF (USA).
We acknowledge the computing resources that are provided by CERN, IN2P3
(France), KIT and DESY (Germany), INFN (Italy), SURF (Netherlands),
PIC (Spain), GridPP (United Kingdom), RRCKI and Yandex
LLC (Russia), CSCS (Switzerland), IFIN-HH (Romania), CBPF (Brazil),
PL-GRID (Poland) and OSC (USA).
We are indebted to the communities behind the multiple open-source
software packages on which we depend.
Individual groups or members have received support from
AvH Foundation (Germany);
EPLANET, Marie Sk\l{}odowska-Curie Actions and ERC (European Union);
ANR, Labex P2IO and OCEVU, and R\'{e}gion Auvergne-Rh\^{o}ne-Alpes (France);
Key Research Program of Frontier Sciences of CAS, CAS PIFI, and the Thousand Talents Program (China);
RFBR, RSF and Yandex LLC (Russia);
GVA, XuntaGal and GENCAT (Spain);
Herchel Smith Fund, the Royal Society, the English-Speaking Union and the Leverhulme Trust (United Kingdom);
Laboratory Directed Research and Development program of LANL (USA).



\addcontentsline{toc}{section}{References}
\setboolean{inbibliography}{true}

\begin{mcitethebibliography}{10}
\mciteSetBstSublistMode{n}
\mciteSetBstMaxWidthForm{subitem}{\alph{mcitesubitemcount})}
\mciteSetBstSublistLabelBeginEnd{\mcitemaxwidthsubitemform\space}
{\relax}{\relax}

\bibitem{PhysRevLett.10.531}
N.~Cabibbo, \ifthenelse{\boolean{articletitles}}{\emph{{Unitary symmetry and
  leptonic decays}},
  }{}\href{https://doi.org/10.1103/PhysRevLett.10.531}{Phys.\ Rev.\ Lett.\
  \textbf{10} (1963) 531}\relax
\mciteBstWouldAddEndPuncttrue
\mciteSetBstMidEndSepPunct{\mcitedefaultmidpunct}
{\mcitedefaultendpunct}{\mcitedefaultseppunct}\relax
\EndOfBibitem
\bibitem{PTP.49.652}
M.~Kobayashi and T.~Maskawa,
  \ifthenelse{\boolean{articletitles}}{\emph{{\CP-violation in the
  renormalizable theory of weak interaction}},
  }{}\href{https://doi.org/10.1143/PTP.49.652}{Progress of Theoretical Physics
  \textbf{49} (1973) 652}\relax
\mciteBstWouldAddEndPuncttrue
\mciteSetBstMidEndSepPunct{\mcitedefaultmidpunct}
{\mcitedefaultendpunct}{\mcitedefaultseppunct}\relax
\EndOfBibitem
\bibitem{Brod:2013sga}
J.~Brod and J.~Zupan, \ifthenelse{\boolean{articletitles}}{\emph{{The ultimate
  theoretical error on $\gamma$ from $B \to DK$ decays}},
  }{}\href{https://doi.org/10.1007/JHEP01(2014)051}{JHEP \textbf{01} (2014)
  051}, \href{http://arxiv.org/abs/1308.5663}{{\normalfont\ttfamily
  arXiv:1308.5663}}\relax
\mciteBstWouldAddEndPuncttrue
\mciteSetBstMidEndSepPunct{\mcitedefaultmidpunct}
{\mcitedefaultendpunct}{\mcitedefaultseppunct}\relax
\EndOfBibitem
\bibitem{Brod:2014bfa}
J.~Brod, A.~Lenz, G.~Tetlalmatzi-Xolocotzi, and M.~Wiebusch,
  \ifthenelse{\boolean{articletitles}}{\emph{{New physics effects in tree-level
  decays and the precision in the determination of the quark mixing angle
  $\gamma$}}, }{}\href{https://doi.org/10.1103/PhysRevD.92.033002}{Phys.\ Rev.\
   \textbf{D92} (2015) 033002},
  \href{http://arxiv.org/abs/1412.1446}{{\normalfont\ttfamily
  arXiv:1412.1446}}\relax
\mciteBstWouldAddEndPuncttrue
\mciteSetBstMidEndSepPunct{\mcitedefaultmidpunct}
{\mcitedefaultendpunct}{\mcitedefaultseppunct}\relax
\EndOfBibitem
\bibitem{CKMfitter2013}
J.~Charles {\em et~al.}, \ifthenelse{\boolean{articletitles}}{\emph{{Future
  sensitivity to new physics in $B_d, B_s$, and K mixings}},
  }{}\href{https://doi.org/10.1103/PhysRevD.89.033016}{Phys.\ Rev.\
  \textbf{D89} (2014) 033016},
  \href{http://arxiv.org/abs/1309.2293}{{\normalfont\ttfamily
  arXiv:1309.2293}}\relax
\mciteBstWouldAddEndPuncttrue
\mciteSetBstMidEndSepPunct{\mcitedefaultmidpunct}
{\mcitedefaultendpunct}{\mcitedefaultseppunct}\relax
\EndOfBibitem
\bibitem{Bevan:2014iga}
\belle and BaBar collaborations, A.~J. Bevan {\em et~al.},
  \ifthenelse{\boolean{articletitles}}{\emph{{The Physics of the B Factories}},
  }{}\href{https://doi.org/10.1140/epjc/s10052-014-3026-9}{Eur.\ Phys.\ J.\
  \textbf{C74} (2014) 3026},
  \href{http://arxiv.org/abs/1406.6311}{{\normalfont\ttfamily
  arXiv:1406.6311}}\relax
\mciteBstWouldAddEndPuncttrue
\mciteSetBstMidEndSepPunct{\mcitedefaultmidpunct}
{\mcitedefaultendpunct}{\mcitedefaultseppunct}\relax
\EndOfBibitem
\bibitem{HFLAV16}
Heavy Flavor Averaging Group, Y.~Amhis {\em et~al.},
  \ifthenelse{\boolean{articletitles}}{\emph{{Averages of $b$-hadron,
  $c$-hadron, and $\tau$-lepton properties as of summer 2016}},
  }{}\href{https://doi.org/10.1140/epjc/s10052-017-5058-4}{Eur.\ Phys.\ J.\
  \textbf{C77} (2017) 895},
  \href{http://arxiv.org/abs/1612.07233}{{\normalfont\ttfamily
  arXiv:1612.07233}}, {updated results and plots available at
  \href{https://hflav.web.cern.ch}{{\texttt{https://hflav.web.cern.ch}}}}\relax
\mciteBstWouldAddEndPuncttrue
\mciteSetBstMidEndSepPunct{\mcitedefaultmidpunct}
{\mcitedefaultendpunct}{\mcitedefaultseppunct}\relax
\EndOfBibitem
\bibitem{LHCb-PAPER-2016-032}
LHCb collaboration, R.~Aaij {\em et~al.},
  \ifthenelse{\boolean{articletitles}}{\emph{{Measurement of the CKM angle
  $\gamma$ from a combination of LHCb results}},
  }{}\href{https://doi.org/10.1007/JHEP12(2016)087}{JHEP \textbf{12} (2016)
  087}, \href{http://arxiv.org/abs/1611.03076}{{\normalfont\ttfamily
  arXiv:1611.03076}}\relax
\mciteBstWouldAddEndPuncttrue
\mciteSetBstMidEndSepPunct{\mcitedefaultmidpunct}
{\mcitedefaultendpunct}{\mcitedefaultseppunct}\relax
\EndOfBibitem
\bibitem{LHCb-CONF-2018-002}
{LHCb collaboration}, \ifthenelse{\boolean{articletitles}}{\emph{{Update of the
  LHCb combination of the CKM angle $\gamma$}}, }{}
  \href{http://cdsweb.cern.ch/search?p=LHCb-CONF-2018-002&f=reportnumber&action_search=Search&c=LHCb+Conference+Contributions}
  {LHCb-CONF-2018-002}\relax
\mciteBstWouldAddEndPuncttrue
\mciteSetBstMidEndSepPunct{\mcitedefaultmidpunct}
{\mcitedefaultendpunct}{\mcitedefaultseppunct}\relax
\EndOfBibitem
\bibitem{LHCb-PAPER-2017-047}
LHCb collaboration, R.~Aaij {\em et~al.},
  \ifthenelse{\boolean{articletitles}}{\emph{{Measurement of \CP asymmetry in
  $\Bs\to D_s^\mp K^\pm$ decays}},
  }{}\href{https://doi.org/10.1007/JHEP03(2018)059}{JHEP \textbf{03} (2018)
  059}, \href{http://arxiv.org/abs/1712.07428}{{\normalfont\ttfamily
  arXiv:1712.07428}}\relax
\mciteBstWouldAddEndPuncttrue
\mciteSetBstMidEndSepPunct{\mcitedefaultmidpunct}
{\mcitedefaultendpunct}{\mcitedefaultseppunct}\relax
\EndOfBibitem
\bibitem{LHCb-PAPER-2018-017}
LHCb collaboration, R.~Aaij {\em et~al.},
  \ifthenelse{\boolean{articletitles}}{\emph{{Measurement of the CKM angle
  $\gamma$ using \decay{\Bpm}{DK^\pm} with \decay{D}{K_S^0 \pi^+ \pi^-,\ \KS
  K^+K^-} decays}},
  }{}\href{http://arxiv.org/abs/1806.01202}{{\normalfont\ttfamily
  arXiv:1806.01202}}, {submitted to JHEP}\relax
\mciteBstWouldAddEndPuncttrue
\mciteSetBstMidEndSepPunct{\mcitedefaultmidpunct}
{\mcitedefaultendpunct}{\mcitedefaultseppunct}\relax
\EndOfBibitem
\bibitem{Gronau:1990ra}
M.~Gronau and D.~London, \ifthenelse{\boolean{articletitles}}{\emph{{How to
  determine all the angles of the unitarity triangle from $\Bd \to D\KS$ and
  $\Bs \to D\phi$}},
  }{}\href{https://doi.org/10.1016/0370-2693(91)91756-L}{Phys.\ Lett.\
  \textbf{B253} (1991) 483}\relax
\mciteBstWouldAddEndPuncttrue
\mciteSetBstMidEndSepPunct{\mcitedefaultmidpunct}
{\mcitedefaultendpunct}{\mcitedefaultseppunct}\relax
\EndOfBibitem
\bibitem{Gronau:2004gt}
M.~Gronau {\em et~al.}, \ifthenelse{\boolean{articletitles}}{\emph{{Using
  untagged $\Bd \to D\KS$ to determine $\gamma$}},
  }{}\href{https://doi.org/10.1103/PhysRevD.69.113003}{Phys.\ Rev.\
  \textbf{D69} (2004) 113003},
  \href{http://arxiv.org/abs/hep-ph/0402055}{{\normalfont\ttfamily
  arXiv:hep-ph/0402055}}\relax
\mciteBstWouldAddEndPuncttrue
\mciteSetBstMidEndSepPunct{\mcitedefaultmidpunct}
{\mcitedefaultendpunct}{\mcitedefaultseppunct}\relax
\EndOfBibitem
\bibitem{Gronau:2007bh}
M.~Gronau, Y.~Grossman, Z.~Surujon, and J.~Zupan,
  \ifthenelse{\boolean{articletitles}}{\emph{{Enhanced effects on extracting
  $\gamma$ from untagged \Bd and \Bs decays}},
  }{}\href{https://doi.org/10.1016/j.physletb.2007.03.057}{Phys.\ Lett.\
  \textbf{B649} (2007) 61},
  \href{http://arxiv.org/abs/hep-ph/0702011}{{\normalfont\ttfamily
  arXiv:hep-ph/0702011}}\relax
\mciteBstWouldAddEndPuncttrue
\mciteSetBstMidEndSepPunct{\mcitedefaultmidpunct}
{\mcitedefaultendpunct}{\mcitedefaultseppunct}\relax
\EndOfBibitem
\bibitem{Ricciardi:1243068}
S.~Ricciardi, \ifthenelse{\boolean{articletitles}}{\emph{{Measuring the CKM
  angle $\gamma$ at LHCb using untagged $B_s \to D \phi$ decays}}, }{}
  \href{http://cdsweb.cern.ch/search?p=LHCb-PUB-2010-005&f=reportnumber&action_search=Search&c=LHCb+Notes}
  {LHCb-PUB-2010-005}\relax
\mciteBstWouldAddEndPuncttrue
\mciteSetBstMidEndSepPunct{\mcitedefaultmidpunct}
{\mcitedefaultendpunct}{\mcitedefaultseppunct}\relax
\EndOfBibitem
\bibitem{Nandi:2011uw}
S.~Nandi and D.~London, \ifthenelse{\boolean{articletitles}}{\emph{{$B_s (\bar
  B_s) \to D^0_{\CP} K {\bar K}$: detecting and discriminating New Physics in
  $B_s$-$\bar B_s$ mixing}},
  }{}\href{https://doi.org/10.1103/PhysRevD.85.114015}{Phys.\ Rev.\
  \textbf{D85} (2012) 114015},
  \href{http://arxiv.org/abs/1108.5769}{{\normalfont\ttfamily
  arXiv:1108.5769}}\relax
\mciteBstWouldAddEndPuncttrue
\mciteSetBstMidEndSepPunct{\mcitedefaultmidpunct}
{\mcitedefaultendpunct}{\mcitedefaultseppunct}\relax
\EndOfBibitem
\bibitem{LHCb-PAPER-2013-035}
LHCb collaboration, R.~Aaij {\em et~al.},
  \ifthenelse{\boolean{articletitles}}{\emph{{Observation of the decay $\Bs\to
  \Dzb\phi$}}, }{}\href{https://doi.org/10.1016/j.physletb.2013.10.057}{Phys.\
  Lett.\  \textbf{B727} (2013) 403},
  \href{http://arxiv.org/abs/1308.4583}{{\normalfont\ttfamily
  arXiv:1308.4583}}\relax
\mciteBstWouldAddEndPuncttrue
\mciteSetBstMidEndSepPunct{\mcitedefaultmidpunct}
{\mcitedefaultendpunct}{\mcitedefaultseppunct}\relax
\EndOfBibitem
\bibitem{LHCb-PAPER-2012-018}
LHCb collaboration, R.~Aaij {\em et~al.},
  \ifthenelse{\boolean{articletitles}}{\emph{{Observation of $\Bz\to
  \Dzb\Kp\Km$ and evidence for $\Bs\to \Dzb\Kp\Km$}},
  }{}\href{https://doi.org/10.1103/PhysRevLett.109.131801}{Phys.\ Rev.\ Lett.\
  \textbf{109} (2012) 131801},
  \href{http://arxiv.org/abs/1207.5991}{{\normalfont\ttfamily
  arXiv:1207.5991}}\relax
\mciteBstWouldAddEndPuncttrue
\mciteSetBstMidEndSepPunct{\mcitedefaultmidpunct}
{\mcitedefaultendpunct}{\mcitedefaultseppunct}\relax
\EndOfBibitem
\bibitem{LHCb-PAPER-2013-022}
LHCb collaboration, R.~Aaij {\em et~al.},
  \ifthenelse{\boolean{articletitles}}{\emph{{Measurements of the branching
  fractions of the decays $\Bs\to \Dzb\Km\pip$ and $\Bz\to \Dzb\Kp\pim$}},
  }{}\href{https://doi.org/10.1103/PhysRevD.87.112009}{Phys.\ Rev.\
  \textbf{D87} (2013) 112009},
  \href{http://arxiv.org/abs/1304.6317}{{\normalfont\ttfamily
  arXiv:1304.6317}}\relax
\mciteBstWouldAddEndPuncttrue
\mciteSetBstMidEndSepPunct{\mcitedefaultmidpunct}
{\mcitedefaultendpunct}{\mcitedefaultseppunct}\relax
\EndOfBibitem
\bibitem{LHCb-PAPER-2013-056}
LHCb collaboration, R.~Aaij {\em et~al.},
  \ifthenelse{\boolean{articletitles}}{\emph{{Study of beauty baryon decays to
  $\Dz\proton h^-$ and $\Lc h^-$ final states}},
  }{}\href{https://doi.org/10.1103/PhysRevD.89.032001}{Phys.\ Rev.\
  \textbf{D89} (2014) 032001},
  \href{http://arxiv.org/abs/1311.4823}{{\normalfont\ttfamily
  arXiv:1311.4823}}\relax
\mciteBstWouldAddEndPuncttrue
\mciteSetBstMidEndSepPunct{\mcitedefaultmidpunct}
{\mcitedefaultendpunct}{\mcitedefaultseppunct}\relax
\EndOfBibitem
\bibitem{LHCb-PAPER-2014-070}
LHCb collaboration, R.~Aaij {\em et~al.},
  \ifthenelse{\boolean{articletitles}}{\emph{{Dalitz plot analysis of $\Bz\to
  \Dzb\pip\pim$ decays}},
  }{}\href{https://doi.org/10.1103/PhysRevD.92.032002}{Phys.\ Rev.\
  \textbf{D92} (2015) 032002},
  \href{http://arxiv.org/abs/1505.01710}{{\normalfont\ttfamily
  arXiv:1505.01710}}\relax
\mciteBstWouldAddEndPuncttrue
\mciteSetBstMidEndSepPunct{\mcitedefaultmidpunct}
{\mcitedefaultendpunct}{\mcitedefaultseppunct}\relax
\EndOfBibitem
\bibitem{LHCb-PAPER-2018-015}
LHCb collaboration, R.~Aaij {\em et~al.},
  \ifthenelse{\boolean{articletitles}}{\emph{{Observation of $\Bs \to \Dstarzb
  \phi$ and search for $\Bd \to \Dzb \phi$ decays}},
  }{}\href{http://arxiv.org/abs/1807.01892}{{\normalfont\ttfamily
  arXiv:1807.01892}}, {submitted to Phys. Rev. Lett.}\relax
\mciteBstWouldAddEndPunctfalse
\mciteSetBstMidEndSepPunct{\mcitedefaultmidpunct}
{}{\mcitedefaultseppunct}\relax
\EndOfBibitem
\bibitem{LHCb-PAPER-2014-036}
LHCb collaboration, R.~Aaij {\em et~al.},
  \ifthenelse{\boolean{articletitles}}{\emph{{Dalitz plot analysis of $\Bs\to
  \Dzb\Km\pip$ decays}},
  }{}\href{https://doi.org/10.1103/PhysRevD.90.072003}{Phys.\ Rev.\
  \textbf{D90} (2014) 072003},
  \href{http://arxiv.org/abs/1407.7712}{{\normalfont\ttfamily
  arXiv:1407.7712}}\relax
\mciteBstWouldAddEndPuncttrue
\mciteSetBstMidEndSepPunct{\mcitedefaultmidpunct}
{\mcitedefaultendpunct}{\mcitedefaultseppunct}\relax
\EndOfBibitem
\bibitem{Alves:2008zz}
LHCb collaboration, A.~A. Alves~Jr.\ {\em et~al.},
  \ifthenelse{\boolean{articletitles}}{\emph{{The \lhcb detector at the LHC}},
  }{}\href{https://doi.org/10.1088/1748-0221/3/08/S08005}{JINST \textbf{3}
  (2008) S08005}\relax
\mciteBstWouldAddEndPuncttrue
\mciteSetBstMidEndSepPunct{\mcitedefaultmidpunct}
{\mcitedefaultendpunct}{\mcitedefaultseppunct}\relax
\EndOfBibitem
\bibitem{LHCb-DP-2014-002}
LHCb collaboration, R.~Aaij {\em et~al.},
  \ifthenelse{\boolean{articletitles}}{\emph{{LHCb detector performance}},
  }{}\href{https://doi.org/10.1142/S0217751X15300227}{Int.\ J.\ Mod.\ Phys.\
  \textbf{A30} (2015) 1530022},
  \href{http://arxiv.org/abs/1412.6352}{{\normalfont\ttfamily
  arXiv:1412.6352}}\relax
\mciteBstWouldAddEndPuncttrue
\mciteSetBstMidEndSepPunct{\mcitedefaultmidpunct}
{\mcitedefaultendpunct}{\mcitedefaultseppunct}\relax
\EndOfBibitem
\bibitem{LHCb-DP-2014-001}
R.~Aaij {\em et~al.}, \ifthenelse{\boolean{articletitles}}{\emph{{Performance
  of the LHCb Vertex Locator}},
  }{}\href{https://doi.org/10.1088/1748-0221/9/09/P09007}{JINST \textbf{9}
  (2014) P09007}, \href{http://arxiv.org/abs/1405.7808}{{\normalfont\ttfamily
  arXiv:1405.7808}}\relax
\mciteBstWouldAddEndPuncttrue
\mciteSetBstMidEndSepPunct{\mcitedefaultmidpunct}
{\mcitedefaultendpunct}{\mcitedefaultseppunct}\relax
\EndOfBibitem
\bibitem{LHCb-DP-2013-003}
R.~Arink {\em et~al.}, \ifthenelse{\boolean{articletitles}}{\emph{{Performance
  of the LHCb Outer Tracker}},
  }{}\href{https://doi.org/10.1088/1748-0221/9/01/P01002}{JINST \textbf{9}
  (2014) P01002}, \href{http://arxiv.org/abs/1311.3893}{{\normalfont\ttfamily
  arXiv:1311.3893}}\relax
\mciteBstWouldAddEndPuncttrue
\mciteSetBstMidEndSepPunct{\mcitedefaultmidpunct}
{\mcitedefaultendpunct}{\mcitedefaultseppunct}\relax
\EndOfBibitem
\bibitem{LHCb-DP-2012-003}
M.~Adinolfi {\em et~al.},
  \ifthenelse{\boolean{articletitles}}{\emph{{Performance of the \lhcb RICH
  detector at the LHC}},
  }{}\href{https://doi.org/10.1140/epjc/s10052-013-2431-9}{Eur.\ Phys.\ J.\
  \textbf{C73} (2013) 2431},
  \href{http://arxiv.org/abs/1211.6759}{{\normalfont\ttfamily
  arXiv:1211.6759}}\relax
\mciteBstWouldAddEndPuncttrue
\mciteSetBstMidEndSepPunct{\mcitedefaultmidpunct}
{\mcitedefaultendpunct}{\mcitedefaultseppunct}\relax
\EndOfBibitem
\bibitem{LHCb-DP-2012-002}
A.~A. Alves~Jr.\ {\em et~al.},
  \ifthenelse{\boolean{articletitles}}{\emph{{Performance of the LHCb muon
  system}}, }{}\href{https://doi.org/10.1088/1748-0221/8/02/P02022}{JINST
  \textbf{8} (2013) P02022},
  \href{http://arxiv.org/abs/1211.1346}{{\normalfont\ttfamily
  arXiv:1211.1346}}\relax
\mciteBstWouldAddEndPuncttrue
\mciteSetBstMidEndSepPunct{\mcitedefaultmidpunct}
{\mcitedefaultendpunct}{\mcitedefaultseppunct}\relax
\EndOfBibitem
\bibitem{BBDT}
V.~V. Gligorov and M.~Williams,
  \ifthenelse{\boolean{articletitles}}{\emph{{Efficient, reliable and fast
  high-level triggering using a bonsai boosted decision tree}},
  }{}\href{https://doi.org/10.1088/1748-0221/8/02/P02013}{JINST \textbf{8}
  (2013) P02013}, \href{http://arxiv.org/abs/1210.6861}{{\normalfont\ttfamily
  arXiv:1210.6861}}\relax
\mciteBstWouldAddEndPuncttrue
\mciteSetBstMidEndSepPunct{\mcitedefaultmidpunct}
{\mcitedefaultendpunct}{\mcitedefaultseppunct}\relax
\EndOfBibitem
\bibitem{Kuzmin:2006mw}
\belle collaboration, A.~Kuzmin {\em et~al.},
  \ifthenelse{\boolean{articletitles}}{\emph{{Study of $\Bzb \to \Dz \pip \pim$
  decays}}, }{}\href{https://doi.org/10.1103/PhysRevD.76.012006}{Phys.\ Rev.\
  \textbf{D76} (2007) 012006},
  \href{http://arxiv.org/abs/hep-ex/0611054}{{\normalfont\ttfamily
  arXiv:hep-ex/0611054}}\relax
\mciteBstWouldAddEndPuncttrue
\mciteSetBstMidEndSepPunct{\mcitedefaultmidpunct}
{\mcitedefaultendpunct}{\mcitedefaultseppunct}\relax
\EndOfBibitem
\bibitem{Sjostrand:2007gs}
T.~Sj\"{o}strand, S.~Mrenna, and P.~Skands,
  \ifthenelse{\boolean{articletitles}}{\emph{{A brief introduction to PYTHIA
  8.1}}, }{}\href{https://doi.org/10.1016/j.cpc.2008.01.036}{Comput.\ Phys.\
  Commun.\  \textbf{178} (2008) 852},
  \href{http://arxiv.org/abs/0710.3820}{{\normalfont\ttfamily
  arXiv:0710.3820}}\relax
\mciteBstWouldAddEndPuncttrue
\mciteSetBstMidEndSepPunct{\mcitedefaultmidpunct}
{\mcitedefaultendpunct}{\mcitedefaultseppunct}\relax
\EndOfBibitem
\bibitem{LHCb-PROC-2010-056}
I.~Belyaev {\em et~al.}, \ifthenelse{\boolean{articletitles}}{\emph{{Handling
  of the generation of primary events in Gauss, the LHCb simulation
  framework}}, }{}\href{https://doi.org/10.1088/1742-6596/331/3/032047}{{J.\
  Phys.\ Conf.\ Ser.\ } \textbf{331} (2011) 032047}\relax
\mciteBstWouldAddEndPuncttrue
\mciteSetBstMidEndSepPunct{\mcitedefaultmidpunct}
{\mcitedefaultendpunct}{\mcitedefaultseppunct}\relax
\EndOfBibitem
\bibitem{Lange:2001uf}
D.~J. Lange, \ifthenelse{\boolean{articletitles}}{\emph{{The EvtGen particle
  decay simulation package}},
  }{}\href{https://doi.org/10.1016/S0168-9002(01)00089-4}{Nucl.\ Instrum.\
  Meth.\  \textbf{A462} (2001) 152}\relax
\mciteBstWouldAddEndPuncttrue
\mciteSetBstMidEndSepPunct{\mcitedefaultmidpunct}
{\mcitedefaultendpunct}{\mcitedefaultseppunct}\relax
\EndOfBibitem
\bibitem{Golonka:2005pn}
P.~Golonka and Z.~Was, \ifthenelse{\boolean{articletitles}}{\emph{{PHOTOS Monte
  Carlo: A precision tool for QED corrections in $Z$ and $W$ decays}},
  }{}\href{https://doi.org/10.1140/epjc/s2005-02396-4}{Eur.\ Phys.\ J.\
  \textbf{C45} (2006) 97},
  \href{http://arxiv.org/abs/hep-ph/0506026}{{\normalfont\ttfamily
  arXiv:hep-ph/0506026}}\relax
\mciteBstWouldAddEndPuncttrue
\mciteSetBstMidEndSepPunct{\mcitedefaultmidpunct}
{\mcitedefaultendpunct}{\mcitedefaultseppunct}\relax
\EndOfBibitem
\bibitem{Allison:2006ve}
Geant4 collaboration, J.~Allison {\em et~al.},
  \ifthenelse{\boolean{articletitles}}{\emph{{Geant4 developments and
  applications}}, }{}\href{https://doi.org/10.1109/TNS.2006.869826}{IEEE
  Trans.\ Nucl.\ Sci.\  \textbf{53} (2006) 270}\relax
\mciteBstWouldAddEndPuncttrue
\mciteSetBstMidEndSepPunct{\mcitedefaultmidpunct}
{\mcitedefaultendpunct}{\mcitedefaultseppunct}\relax
\EndOfBibitem
\bibitem{Agostinelli:2002hh}
Geant4 collaboration, S.~Agostinelli {\em et~al.},
  \ifthenelse{\boolean{articletitles}}{\emph{{Geant4: A simulation toolkit}},
  }{}\href{https://doi.org/10.1016/S0168-9002(03)01368-8}{Nucl.\ Instrum.\
  Meth.\  \textbf{A506} (2003) 250}\relax
\mciteBstWouldAddEndPuncttrue
\mciteSetBstMidEndSepPunct{\mcitedefaultmidpunct}
{\mcitedefaultendpunct}{\mcitedefaultseppunct}\relax
\EndOfBibitem
\bibitem{LHCb-PROC-2011-006}
M.~Clemencic {\em et~al.}, \ifthenelse{\boolean{articletitles}}{\emph{{The
  \lhcb simulation application, Gauss: Design, evolution and experience}},
  }{}\href{https://doi.org/10.1088/1742-6596/331/3/032023}{{J.\ Phys.\ Conf.\
  Ser.\ } \textbf{331} (2011) 032023}\relax
\mciteBstWouldAddEndPuncttrue
\mciteSetBstMidEndSepPunct{\mcitedefaultmidpunct}
{\mcitedefaultendpunct}{\mcitedefaultseppunct}\relax
\EndOfBibitem
\bibitem{LHCb-PAPER-2012-056}
LHCb collaboration, R.~Aaij {\em et~al.},
  \ifthenelse{\boolean{articletitles}}{\emph{{Search for the decay $\Bs\to
  \Dssmp\pipm$}}, }{}\href{https://doi.org/10.1103/PhysRevD.87.071101}{Phys.\
  Rev.\  \textbf{D87} (2013) 071101(R)},
  \href{http://arxiv.org/abs/1302.6446}{{\normalfont\ttfamily
  arXiv:1302.6446}}\relax
\mciteBstWouldAddEndPuncttrue
\mciteSetBstMidEndSepPunct{\mcitedefaultmidpunct}
{\mcitedefaultendpunct}{\mcitedefaultseppunct}\relax
\EndOfBibitem
\bibitem{PDG2018}
Particle Data Group, M.~Tanabashi {\em et~al.},
  \ifthenelse{\boolean{articletitles}}{\emph{{\href{http://pdg.lbl.gov/}{Review
  of particle physics}}}, }{}Phys.\ Rev.\  \textbf{D98} (2018) 030001\relax
\mciteBstWouldAddEndPuncttrue
\mciteSetBstMidEndSepPunct{\mcitedefaultmidpunct}
{\mcitedefaultendpunct}{\mcitedefaultseppunct}\relax
\EndOfBibitem
\bibitem{Hulsbergen:2005pu}
W.~D. Hulsbergen, \ifthenelse{\boolean{articletitles}}{\emph{{Decay chain
  fitting with a Kalman filter}},
  }{}\href{https://doi.org/10.1016/j.nima.2005.06.078}{Nucl.\ Instrum.\ Meth.\
  \textbf{A552} (2005) 566},
  \href{http://arxiv.org/abs/physics/0503191}{{\normalfont\ttfamily
  arXiv:physics/0503191}}\relax
\mciteBstWouldAddEndPuncttrue
\mciteSetBstMidEndSepPunct{\mcitedefaultmidpunct}
{\mcitedefaultendpunct}{\mcitedefaultseppunct}\relax
\EndOfBibitem
\bibitem{Hocker:2007ht}
H.~Voss, A.~Hoecker, J.~Stelzer, and F.~Tegenfeldt,
  \ifthenelse{\boolean{articletitles}}{\emph{{TMVA - Toolkit for Multivariate
  Data Analysis}}, }{}\href{https://doi.org/10.22323/1.050.0040}{PoS
  \textbf{ACAT} (2007) 040}\relax
\mciteBstWouldAddEndPuncttrue
\mciteSetBstMidEndSepPunct{\mcitedefaultmidpunct}
{\mcitedefaultendpunct}{\mcitedefaultseppunct}\relax
\EndOfBibitem
\bibitem{TMVA4}
A.~Hoecker {\em et~al.}, \ifthenelse{\boolean{articletitles}}{\emph{{TMVA 4 ---
  Toolkit for Multivariate Data Analysis. Users Guide.}},
  }{}\href{http://arxiv.org/abs/physics/0703039}{{\normalfont\ttfamily
  arXiv:physics/0703039}}\relax
\mciteBstWouldAddEndPuncttrue
\mciteSetBstMidEndSepPunct{\mcitedefaultmidpunct}
{\mcitedefaultendpunct}{\mcitedefaultseppunct}\relax
\EndOfBibitem
\bibitem{Pivk:2004ty}
M.~Pivk and F.~R. Le~Diberder,
  \ifthenelse{\boolean{articletitles}}{\emph{{sPlot: a statistical tool to
  unfold data distributions}},
  }{}\href{https://doi.org/10.1016/j.nima.2005.08.106}{Nucl.\ Instrum.\ Meth.\
  \textbf{A555} (2005) 356},
  \href{http://arxiv.org/abs/physics/0402083}{{\normalfont\ttfamily
  arXiv:physics/0402083}}\relax
\mciteBstWouldAddEndPuncttrue
\mciteSetBstMidEndSepPunct{\mcitedefaultmidpunct}
{\mcitedefaultendpunct}{\mcitedefaultseppunct}\relax
\EndOfBibitem
\bibitem{CBFunction}
T.~Skwarnicki, {\em {A study of the radiative CASCADE transitions between the
  Upsilon-Prime and Upsilon resonances}}, PhD thesis, Institute of Nuclear
  Physics, Krakow, 1986,
  {\href{http://inspirehep.net/record/230779/}{DESY-F31-86-02}}\relax
\mciteBstWouldAddEndPuncttrue
\mciteSetBstMidEndSepPunct{\mcitedefaultmidpunct}
{\mcitedefaultendpunct}{\mcitedefaultseppunct}\relax
\EndOfBibitem
\bibitem{Fisher:1936et}
R.~A. Fisher, \ifthenelse{\boolean{articletitles}}{\emph{{The use of multiple
  measurements in taxonomic problems}},
  }{}\href{https://doi.org/10.1111/j.1469-1809.1936.tb02137.x}{Annals Eugen.\
  \textbf{7} (1936) 179}\relax
\mciteBstWouldAddEndPuncttrue
\mciteSetBstMidEndSepPunct{\mcitedefaultmidpunct}
{\mcitedefaultendpunct}{\mcitedefaultseppunct}\relax
\EndOfBibitem
\bibitem{Koppenburg:2017zsh}
P.~Koppenburg, \ifthenelse{\boolean{articletitles}}{\emph{{Statistical biases
  in measurements with multiple candidates}},
  }{}\href{http://arxiv.org/abs/1703.01128}{{\normalfont\ttfamily
  arXiv:1703.01128}}\relax
\mciteBstWouldAddEndPuncttrue
\mciteSetBstMidEndSepPunct{\mcitedefaultmidpunct}
{\mcitedefaultendpunct}{\mcitedefaultseppunct}\relax
\EndOfBibitem
\bibitem{Zupanc:2013iki}
\belle collaboration, A.~Zupanc {\em et~al.},
  \ifthenelse{\boolean{articletitles}}{\emph{{Measurement of the Branching
  Fraction $\mathcal B(\Lc \to p K^- \pi^+)$}},
  }{}\href{https://doi.org/10.1103/PhysRevLett.113.042002}{Phys.\ Rev.\ Lett.\
  \textbf{113} (2014) 042002},
  \href{http://arxiv.org/abs/1312.7826}{{\normalfont\ttfamily
  arXiv:1312.7826}}\relax
\mciteBstWouldAddEndPuncttrue
\mciteSetBstMidEndSepPunct{\mcitedefaultmidpunct}
{\mcitedefaultendpunct}{\mcitedefaultseppunct}\relax
\EndOfBibitem
\bibitem{fsfd}
LHCb collaboration, R.~Aaij {\em et~al.},
  \ifthenelse{\boolean{articletitles}}{\emph{{Measurement of the fragmentation
  fraction ratio $f_s/f_d$ and its dependence on $B$ meson kinematics}},
  }{}\href{https://doi.org/10.1007/JHEP04(2013)001}{JHEP \textbf{04} (2013)
  001}, \href{http://arxiv.org/abs/1301.5286}{{\normalfont\ttfamily
  arXiv:1301.5286}}, $f_s/f_d$ value updated in
  \href{https://cds.cern.ch/record/1559262}{LHCb-CONF-2013-011}\relax
\mciteBstWouldAddEndPuncttrue
\mciteSetBstMidEndSepPunct{\mcitedefaultmidpunct}
{\mcitedefaultendpunct}{\mcitedefaultseppunct}\relax
\EndOfBibitem
\bibitem{LHCb-PAPER-2014-004}
LHCb collaboration, R.~Aaij {\em et~al.},
  \ifthenelse{\boolean{articletitles}}{\emph{{Study of the kinematic
  dependences of $\Lb$ production in $\proton\proton$ collisions and a
  measurement of the $\Lb\to \Lc\pim$ branching fraction}},
  }{}\href{https://doi.org/10.1007/JHEP08(2014)143}{JHEP \textbf{08} (2014)
  143}, \href{http://arxiv.org/abs/1405.6842}{{\normalfont\ttfamily
  arXiv:1405.6842}}\relax
\mciteBstWouldAddEndPuncttrue
\mciteSetBstMidEndSepPunct{\mcitedefaultmidpunct}
{\mcitedefaultendpunct}{\mcitedefaultseppunct}\relax
\EndOfBibitem
\bibitem{Satpathy:2002js}
\belle collaboration, A.~Satpathy {\em et~al.},
  \ifthenelse{\boolean{articletitles}}{\emph{{Study of $\Bzb \to \Dstarz \pip
  \pim$ decays}},
  }{}\href{https://doi.org/10.1016/S0370-2693(02)03198-2}{Phys.\ Lett.\
  \textbf{B553} (2003) 159},
  \href{http://arxiv.org/abs/hep-ex/0211022}{{\normalfont\ttfamily
  arXiv:hep-ex/0211022}}\relax
\mciteBstWouldAddEndPuncttrue
\mciteSetBstMidEndSepPunct{\mcitedefaultmidpunct}
{\mcitedefaultendpunct}{\mcitedefaultseppunct}\relax
\EndOfBibitem
\bibitem{Cranmer:2000du}
K.~S. Cranmer, \ifthenelse{\boolean{articletitles}}{\emph{{Kernel estimation in
  high-energy physics}},
  }{}\href{https://doi.org/10.1016/S0010-4655(00)00243-5}{Comput.\ Phys.\
  Commun.\  \textbf{136} (2001) 198},
  \href{http://arxiv.org/abs/hep-ex/0011057}{{\normalfont\ttfamily
  arXiv:hep-ex/0011057}}\relax
\mciteBstWouldAddEndPuncttrue
\mciteSetBstMidEndSepPunct{\mcitedefaultmidpunct}
{\mcitedefaultendpunct}{\mcitedefaultseppunct}\relax
\EndOfBibitem
\bibitem{James:2004xla}
F.~James, \ifthenelse{\boolean{articletitles}}{\emph{{MINUIT - Function
  minimization and error analysis}}, }{} 1994.
\newblock {\href{http://hep.fi.infn.it/minuit.pdf}{CERN Program Library Long
  Writeup D506}}\relax
\mciteBstWouldAddEndPuncttrue
\mciteSetBstMidEndSepPunct{\mcitedefaultmidpunct}
{\mcitedefaultendpunct}{\mcitedefaultseppunct}\relax
\EndOfBibitem
\bibitem{Verkerke:2003ir}
W.~Verkerke and D.~P. Kirkby, \ifthenelse{\boolean{articletitles}}{\emph{{The
  RooFit toolkit for data modeling}}, }{}eConf \textbf{C0303241} (2003)
  MOLT007, \href{http://arxiv.org/abs/physics/0306116}{{\normalfont\ttfamily
  arXiv:physics/0306116}}\relax
\mciteBstWouldAddEndPuncttrue
\mciteSetBstMidEndSepPunct{\mcitedefaultmidpunct}
{\mcitedefaultendpunct}{\mcitedefaultseppunct}\relax
\EndOfBibitem
\bibitem{LHCb:12113055}
R.~Aaij {\em et~al.}, \ifthenelse{\boolean{articletitles}}{\emph{{The LHCb
  Trigger and its Performance in 2011}},
  }{}\href{https://doi.org/10.1088/1748-0221/8/04/P04022}{JINST \textbf{8}
  (2013) P04022}, \href{http://arxiv.org/abs/1211.3055}{{\normalfont\ttfamily
  arXiv:1211.3055}}\relax
\mciteBstWouldAddEndPuncttrue
\mciteSetBstMidEndSepPunct{\mcitedefaultmidpunct}
{\mcitedefaultendpunct}{\mcitedefaultseppunct}\relax
\EndOfBibitem
\bibitem{MartinSanchez:2012yxa}
A.~Martin~Sanchez, P.~Robbe, and M.-H. Schune,
  \ifthenelse{\boolean{articletitles}}{\emph{{Performances of the LHCb L0
  Calorimeter Trigger}}, }{}
  \href{http://cdsweb.cern.ch/search?p=LHCb-PUB-2011-026&f=reportnumber&action_search=Search&c=LHCb+Notes}
  {LHCb-PUB-2011-026}\relax
\mciteBstWouldAddEndPuncttrue
\mciteSetBstMidEndSepPunct{\mcitedefaultmidpunct}
{\mcitedefaultendpunct}{\mcitedefaultseppunct}\relax
\EndOfBibitem
\bibitem{LHCb-PAPER-2014-028}
LHCb collaboration, R.~Aaij {\em et~al.},
  \ifthenelse{\boolean{articletitles}}{\emph{{Measurement of \CP violation
  parameters in $\Bz\to \D\Kstarz$ decays}},
  }{}\href{https://doi.org/10.1103/PhysRevD.90.112002}{Phys.\ Rev.\
  \textbf{D90} (2014) 112002},
  \href{http://arxiv.org/abs/1407.8136}{{\normalfont\ttfamily
  arXiv:1407.8136}}\relax
\mciteBstWouldAddEndPuncttrue
\mciteSetBstMidEndSepPunct{\mcitedefaultmidpunct}
{\mcitedefaultendpunct}{\mcitedefaultseppunct}\relax
\EndOfBibitem
\bibitem{LHCb-PAPER-2011-022}
LHCb collaboration, R.~Aaij {\em et~al.},
  \ifthenelse{\boolean{articletitles}}{\emph{{Measurements of the branching
  fractions of the decays $\Bs\to \Dsmp\Kpm$ and $\Bs\to \Dsm\pip$}},
  }{}\href{https://doi.org/10.1007/JHEP06(2012)115}{JHEP \textbf{06} (2012)
  115}, \href{http://arxiv.org/abs/1204.1237}{{\normalfont\ttfamily
  arXiv:1204.1237}}\relax
\mciteBstWouldAddEndPuncttrue
\mciteSetBstMidEndSepPunct{\mcitedefaultmidpunct}
{\mcitedefaultendpunct}{\mcitedefaultseppunct}\relax
\EndOfBibitem
\bibitem{Lees:2011gw}
\babar collaboration, J.~P. Lees {\em et~al.},
  \ifthenelse{\boolean{articletitles}}{\emph{{Branching fraction measurements
  of the color-suppressed decays $\bar{B}^0 \to D^{(*)0} \pi^0$, $D^{(*)0}
  \eta$, $D^{(*)0} \omega$, and $D^{(*)0} \eta^\prime$ and measurement of the
  polarization in the decay $\bar{B}^0 \to D^{*0} \omega$}},
  }{}\href{https://doi.org/10.1103/PhysRevD.84.112007}{Phys.\ Rev.\
  \textbf{D84} (2011) 112007},
  \href{http://arxiv.org/abs/1107.5751}{{\normalfont\ttfamily
  arXiv:1107.5751}},
  {\href{https://journals.aps.org/prd/abstract/10.1103/PhysRevD.87.039901}{[Erratum:
  Phys. Rev.D 87, 039901(2013)]}}\relax
\mciteBstWouldAddEndPuncttrue
\mciteSetBstMidEndSepPunct{\mcitedefaultmidpunct}
{\mcitedefaultendpunct}{\mcitedefaultseppunct}\relax
\EndOfBibitem
\bibitem{Matvienko:2015gqa}
\belle collaboration, D.~Matvienko {\em et~al.},
  \ifthenelse{\boolean{articletitles}}{\emph{{Study of $D^{**}$ production and
  light hadronic states in the $\bar{B}^0 \to D^{*+} \omega \pi^-$ decay}},
  }{}\href{https://doi.org/10.1103/PhysRevD.92.012013}{Phys.\ Rev.\
  \textbf{D92} (2015) 012013},
  \href{http://arxiv.org/abs/1505.03362}{{\normalfont\ttfamily
  arXiv:1505.03362}}\relax
\mciteBstWouldAddEndPuncttrue
\mciteSetBstMidEndSepPunct{\mcitedefaultmidpunct}
{\mcitedefaultendpunct}{\mcitedefaultseppunct}\relax
\EndOfBibitem
\bibitem{Csorna:2003bw}
\cleo collaboration, S.~E. Csorna {\em et~al.},
  \ifthenelse{\boolean{articletitles}}{\emph{{Measurements of the branching
  fractions and helicity amplitudes in $B \to D^{*} \rho$ decays}},
  }{}\href{https://doi.org/10.1103/PhysRevD.67.112002}{Phys.\ Rev.\
  \textbf{D67} (2003) 112002},
  \href{http://arxiv.org/abs/hep-ex/0301028}{{\normalfont\ttfamily
  arXiv:hep-ex/0301028}}\relax
\mciteBstWouldAddEndPuncttrue
\mciteSetBstMidEndSepPunct{\mcitedefaultmidpunct}
{\mcitedefaultendpunct}{\mcitedefaultseppunct}\relax
\EndOfBibitem
\bibitem{Aubert:2003ae}
\babar collaboration, B.~Aubert {\em et~al.},
  \ifthenelse{\boolean{articletitles}}{\emph{{Measurement of the branching
  fraction and polarization for the decay $B^- \to D^{0*} K^{*-}$}},
  }{}\href{https://doi.org/10.1103/PhysRevLett.92.141801}{Phys.\ Rev.\ Lett.\
  \textbf{92} (2004) 141801},
  \href{http://arxiv.org/abs/hep-ex/0308057}{{\normalfont\ttfamily
  arXiv:hep-ex/0308057}}\relax
\mciteBstWouldAddEndPuncttrue
\mciteSetBstMidEndSepPunct{\mcitedefaultmidpunct}
{\mcitedefaultendpunct}{\mcitedefaultseppunct}\relax
\EndOfBibitem
\bibitem{LHCb-PAPER-2014-035}
LHCb collaboration, R.~Aaij {\em et~al.},
  \ifthenelse{\boolean{articletitles}}{\emph{{Observation of overlapping
  spin-$1$ and spin-$3$ $\Dzb\Km$ resonances at mass $2.86$\gevcc}},
  }{}\href{https://doi.org/10.1103/PhysRevLett.113.162001}{Phys.\ Rev.\ Lett.\
  \textbf{113} (2014) 162001},
  \href{http://arxiv.org/abs/1407.7574}{{\normalfont\ttfamily
  arXiv:1407.7574}}\relax
\mciteBstWouldAddEndPuncttrue
\mciteSetBstMidEndSepPunct{\mcitedefaultmidpunct}
{\mcitedefaultendpunct}{\mcitedefaultseppunct}\relax
\EndOfBibitem
\end{mcitethebibliography}
\ifx\mcitethebibliography\mciteundefinedmacro
\PackageError{LHCb.bst}{mciteplus.sty has not been loaded}
{This bibstyle requires the use of the mciteplus package.}\fi
\providecommand{\href}[2]{#2}

\newpage

\newpage
\centerline{\large\bf LHCb collaboration}
\begin{flushleft}
\small
R.~Aaij$^{27}$,
B.~Adeva$^{41}$,
M.~Adinolfi$^{48}$,
C.A.~Aidala$^{73}$,
Z.~Ajaltouni$^{5}$,
S.~Akar$^{59}$,
P.~Albicocco$^{18}$,
J.~Albrecht$^{10}$,
F.~Alessio$^{42}$,
M.~Alexander$^{53}$
% ...

{\footnotesize \it
% ...
$ ^{5}$Clermont Universit{\'e}, Universit{\'e} Blaise Pascal, CNRS/IN2P3, LPC, Clermont-Ferrand, France\\
% ...
$ ^{10}$Fakult{\"a}t Physik, Technische Universit{\"a}t Dortmund, Dortmund, Germany\\
% ...
$ ^{18}$INFN Laboratori Nazionali di Frascati, Frascati, Italy\\
% ...
$ ^{27}$Nikhef National Institute for Subatomic Physics, Amsterdam, Netherlands\\
% ...
$ ^{41}$Instituto Galego de F{\'\i}sica de Altas Enerx{\'\i}as (IGFAE), Universidade de Santiago de Compostela, Santiago de Compostela, Spain\\
$ ^{42}$European Organization for Nuclear Research (CERN), Geneva, Switzerland\\
% ...
$ ^{48}$H.H. Wills Physics Laboratory, University of Bristol, Bristol, United Kingdom\\
% ...
$ ^{53}$School of Physics and Astronomy, University of Glasgow, Glasgow, United Kingdom\\
% ...
$ ^{59}$University of Cincinnati, Cincinnati, OH, United States\\
% ...
$ ^{61}$Syracuse University, Syracuse, NY, United States\\
% ...
$ ^{73}$University of Michigan, Ann Arbor, United States, associated to $^{61}$\\
% ...
}
\end{flushleft}

\end{document}